\documentclass[aps,pra,twocolumn,superscriptaddress,10pt,article,showpacs,longbibliography]{revtex4-2}
\usepackage{blindtext}
\usepackage{lipsum}
\usepackage{graphics}
\usepackage{amsmath}
\usepackage{graphicx}
\usepackage{graphics}
\usepackage{amssymb}
\usepackage{verbatim}
\usepackage{physics}
\usepackage{float}
\usepackage{bm}
\usepackage[normalem]{ulem}
\usepackage[colorlinks=true, citecolor=dodgerblue,linkcolor=dodgerblue,urlcolor=dodgerblue,pdftitle={Polar Sensing}]{hyperref}
\newcommand{\bo}[1]{\mathbf{#1}}
\newcommand{\phii}{\varphi}

\def \r{{\mathbf r}}
\def \rp{{\mathbf r^\prime}}

\def \q{{\mathbf q}}

\def \r{{\mathbf r}}
\def \E{{\mathbf E}}
\def \P{{\mathbf P}}

\def \p{{\mathbf p}}
\def \v{{\mathbf v}}

\def \beq{\begin{eqnarray}}
\def \eeq{\end{eqnarray}}

\usepackage[dvipsnames]{xcolor}
\def \q{{\mathbf q}}
\def \nn{\nonumber \\}

\def \beq{\begin{eqnarray}}
\def \eeq{\end{eqnarray}}
\def \nn{\nonumber \\}

\definecolor{darkorange}{HTML}{FF8C00}
\definecolor{orange(ryb)}{HTML}{FFA500}
\definecolor{dodgerblue}{HTML}{1E90FF}
\definecolor{pinkerton}{HTML}{EC368D}
\definecolor{forest}{HTML}{6DD189}

\newcommand{\rss}[1]{\textcolor{dodgerblue}{#1}}

\begin{document}
\title{Noise Electrometry of Polar and Dielectric Materials}
\author{Rahul Sahay}
\affiliation{Department of Physics, Harvard University, Cambridge, Massachusetts 02138, USA}
\affiliation{Department of Physics, University of California, Berkeley, California 94720, USA}
\author{Pavel A. Volkov}
\affiliation{
Department of Physics, University of Connecticut, Storrs, Connecticut 06269, USA
}
\affiliation{Department of Physics, Harvard University, Cambridge, Massachusetts 02138, USA}
\author{Satcher Hsieh}
\affiliation{Department of Physics, University of California, Berkeley, California 94720, USA}
\affiliation{Materials Sciences Division, Lawrence Berkeley National Laboratory, Berkeley, California 94720, USA}
\author{Eric Parsonnet}
\affiliation{Department of Physics, University of California, Berkeley, California 94720, USA}
\author{Lane W. Martin}
\affiliation{Departments of Materials Science and NanoEngineering, Chemistry, and Physics and Astronomy, Rice University, Houston, TX 77005, USA}
\affiliation{Rice Advanced Materials Institute, Rice University, Houston, TX 77005, USA}
\author{Ramamoorthy Ramesh}
\affiliation{Department of Physics, University of California, Berkeley, California 94720, USA}
\affiliation{Materials Sciences Division, Lawrence Berkeley National Laboratory, Berkeley, California 94720, USA}
\affiliation{Department of Materials Science and Engineering, University of California, Berkeley, California 94720, USA}
\author{Norman Y. Yao}
\affiliation{Department of Physics, Harvard University, Cambridge, Massachusetts 02138, USA}
\affiliation{Department of Physics, University of California, Berkeley, California 94720, USA}
\affiliation{Materials Sciences Division, Lawrence Berkeley National Laboratory, Berkeley, California 94720, USA}
\author{Shubhayu Chatterjee}
\affiliation{Department of Physics, University of California, Berkeley, California 94720, USA}
\affiliation{Department of Physics, Carnegie Mellon University, Pittsburgh, PA 15213, USA}

\begin{abstract}
A qubit sensor with an electric dipole moment acquires an additional contribution to its depolarization rate when it is placed in the vicinity of a polar or dielectric material as a consequence of electrical noise arising from polarization fluctuations in the latter. 
Here, we characterize this relaxation rate as a function of experimentally tunable parameters such as sample-probe distance, probe-frequency, and temperature, and demonstrate that it offers a window into dielectric properties of insulating materials over a wide range of frequencies and length scales.
We discuss the experimental feasibility of our proposal and illustrate its ability to probe a variety of phenomena, ranging from collective polar excitations to phase transitions and disorder-dominated physics in relaxor ferroelectrics.
Our proposal paves the way for a novel table-top probe of polar and dielectric materials in a parameter regime complementary to existing tools and techniques.

\end{abstract}

\maketitle
\begin{figure}[!t]
    \centering
    \includegraphics[width = 247pt]{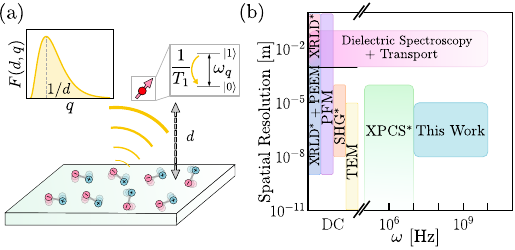}
\caption{\textbf{Overview of Qubit Noise Electrometry.} (a) We propose a qubit sensing experiment where a probe qubit (top right), with splitting $\omega_q$, is a distance $d$ away from a material. Fluctuations in the material's dipoles source electrical noise at the qubit causing it to relax from $\ket{1}$ to $\ket{0}$ at a rate $1/T_1$. The qubit is sensitive to fluctuations at frequency $\omega_q$ and wavevectors near $1/d$ (see filter on top left). In panel (b), we show the regimes of applicability of qubit sensors and other probes including microscopy techniques [atomic-force,  piezoresponse-force, and transmission electron microscopy (AFM, PFM, and TEM)], spectroscopy techniques [x-ray photon correlation, second harmonic generation, x-ray linear dichroism spectroscopy with and without photoemission electron microscopy (XPCS, SHG, XRLD, PEEM)] and electrical transport techniques~\cite{SHGReview_McCullian2020, XRLD_Polisetty_2012, PFM_Review_Gruverman2019, XPCS_Grbel2008, TEM_Winkler2012, DielectricSpectroscopy_Grigas2019, Parsonnet2020, pump_probe_ft}. Techniques often requiring high intensity light sources are marked with a $*$. }
    \label{fig:Fig_1}
\end{figure}

\textbf{Introduction.} Polar and dielectric materials exhibit a plethora of interesting correlated physics~\cite{RameshSkyrmions, Vlad, Volkov, Balatsky} and are emerging as key components in next-generation solid-state technologies~\cite{ncfet, feram, meso, nikonov, Martin2016}.
Consequently, a multitude of techniques for probing them have been developed, ranging from different forms of microscopy and spectroscopy to electrical transport (Fig.~\ref{fig:Fig_1}b)~\cite{SHGReview_McCullian2020, XRLD_Polisetty_2012, PFM_Review_Gruverman2019, XPCS_Grbel2008, TEM_Winkler2012, DielectricSpectroscopy_Grigas2019, Parsonnet2020}.
While these methods have led to incredible scientific progress,  several outstanding questions, such as the origin of polar instabilities in ultra-thin ferroelectric films~\cite{Martin2016} and the structure of polar domains in relaxor ferroelectrics~\cite{takenaka2017slush}, remain formidable challenges.
In part, this is due to the difficulty of probing the near-equilibrium polar dynamics of thin samples over a wide range of length and time scales simultaneously~\cite{Martin2016}---which at present requires the use of high-intensity synchrotron light sources.
As such, developing a table-top probe with the requisite frequency and spatial resolution would naturally complement existing experimental probes of polar and dielectric materials.

Nanoscale quantum sensors, typically based on atomic-scale impurities embedded in insulating materials, provide a candidate for such a probe.
Such sensors can be excellent AC electrometers and magnetometers: they can probe a wide frequency range and locally image both static configurations and dynamic fluctuations of electromagnetic fields with nanoscale resolution~\cite{Degen_RMP,nvsinglespin,Dovzhenko2018,Satcher, Block2020, Mittiga2018}.
As such, numerous theoretical proposals and pioneering experiments have utilized their magnetic sensing capabilities to probe spin dynamics and current fluctuations in solid-state systems~\cite{Kolkowitz,Casola,Agarwal2017,Joaquin18,Finco2021,Schafer2014,Li2019,Schmid2015,Ariyaratne2018,Steinert2013,Ermakova2013, MagnonScatteringPlatform, mccullian_broadband_2020, Joaquin_magnon_18, thermalmagnon_NV,Rustagi,Zhang2020, CRD2018,CD2021,DC2021,Flebus2018,Flebus2019,McLaughlin2021,Satcher,SC_PT,dolgirev2024local,wang2022noninvasive,xu2020strain,curtis2024probing,ziffer2024quantum,melendez2025quantum,liu2025quantum}.

In this Letter, we show that the \textit{electrical} sensing capabilities of single-qubit sensors can be used to probe the physics of polar and dielectric materials, even in the thin-film context.
In particular, we demonstrate that the relaxation rate of a qubit in the presence of electrical noise arising from such materials encodes the material's dielectric properties at frequencies set by the energy splitting of the qubit and wave vectors set by the qubit-sample distance.
Hence, by tuning these two parameters, such qubit sensors can non-invasively and wirelessly probe polar and dielectric materials on frequency scales between $10$~$\text{ MHz}-10~\text{GHz}$ down to nanometer length scales and over a wide range of temperatures, $1$ K - $600$ K \cite{Astner2018,Awschalom2012,Liu2019}.
To highlight the utility of these sensors, we demonstrate how they can (i) detect the presence of exotic collective excitations in polar fluids, (ii) characterize paraelectric-to-ferroelectric phase transitions 
, and (iii) probe local polar dynamics in relaxor ferroelectrics.
Finally, we illustrate the feasibility of qubit sensing via numerical estimates for the relaxation rate signal of a nitrogen-vacancy (NV) center in diamond placed near candidate polar materials.

\textbf{Qubit Relaxometry Concept.} Our experimental proposal is depicted schematically in Fig.~\ref{fig:Fig_1}(a) wherein we envision an isolated qubit sensor placed a distance $d$ away from a polar or dielectric material.
This qubit is a two-level system with a ground state $\ket{0}$ split in energy from an excited state $\ket{1}$ by $\hbar \omega_q$ and its quantum state can be initialized, measured, and manipulated optically.
Moreover, we consider qubits with an electric dipole moment $\hat{\mathbf{d}} = d_{\perp} (\sigma_x \hat{x} + \sigma_y \hat{y})$  and a magnetic moment $\hat{\boldsymbol{\mu}} = \mu_z \sigma_z \hat{z}$, where $\boldsymbol{\sigma}$ are the Pauli matrices, which specifies their coupling to electromagnetic fields as $H_{\text{q-EM}} =\hat{\mathbf{d}} \cdot \mathbf{E} +  \hat{\boldsymbol{\mu}} \cdot \mathbf{B}$.
As a result, electric fields $\mathbf{E}$ drive transitions between $\ket{0}$ and $\ket{1}$ and magnetic fields $\mathbf{B}$ can control their splitting. 

Electrical noise emanating from the material will couple the two states of the qubit and cause the qubit, initialized in its excited state, to relax to a thermal equilibrium set by the ambient temperature, $T$.
The relaxation rate $1/T_1$ is computed from Fermi's Golden Rule to be:
\beq\label{eq-1/T1_EEresonant}
\frac{1}{T_1} = \frac{d_{\perp}^2}{2}\coth\left( \frac{\beta \omega_q}{2} \right) \int_{-\infty}^{\infty} dt\ \langle [E_-(t), E_+(0)] \rangle e^{i \omega_q t}~~~~
\eeq
where the electrical noise is quantified via the auto-correlation function $\langle [E_-(t), E_+(0)] \rangle$ with $E_{\pm} = E_x \pm i E_y$, $\beta = 1/k_B T$,  and $\langle \cdots \rangle$ denotes a thermal average.
Eq.~\eqref{eq-1/T1_EEresonant} expresses that only electrical noise at the qubit-splitting frequency contributes to its relaxation rate.

To understand how the relaxation rate above is connected to the underlying material's dielectric properties, we note that electrical noise in an insulator arises due to thermal or quantum fluctuations of the polarization density, $\mathbf{P}$.
The fluctuations at frequency, $\omega$, and wavevector, $\q$, can be quantified by the retarded polarization correlation function $\chi_{\alpha \beta}(\omega, \q)  = i \int_0^{\infty} dt \, e^{i \omega t} \langle [P_\alpha^{\dagger}(t, \q), P_\beta(0, \q)] \rangle$ ($\alpha, \beta = x, y, z$) which determines the dielectric tensor of the material, $\varepsilon_{\alpha \beta}(\omega, \q) = \delta_{\alpha \beta} + \chi_{\alpha\beta}(\omega, \q)/\varepsilon_0$, and thus encodes its electrical response ~\cite{ft1,hansenMcDonaldBook,grayGubbins}.
%
By utilizing these correlation functions, we can formalize the relationship between fluctuations of polarization in the material and electrical noise at the qubit.
For simplicity, we assume that the material is a stack of $N$, weakly inter-correlated, two-dimensional (2D) monolayers spaced apart by a distance $w$ (modeling a thin-film) and is both translationally and rotationally invariant (see the supplemental for generalizations)~\cite{SM}.
From Maxwell's equations, polarization fluctuations of this sample propagate to electrical noise as:
\beq \label{eq-EE_from_C}
\langle [E_- (t), E_+(0)] \rangle  = \mu_0^2 \int \frac{d\omega\ d^2\q}{(2\pi)^3}\ F(d, q) \mathcal{C}(\omega, \q) e^{-i \omega t}~~~
\eeq
where $\mathcal{C}(\omega, \q) = \text{Im} \left[\chi_{+-}(\omega, \q) + \chi_{-+}(\omega, \q)  + 4\chi_{zz}(\omega, \q) \right]$, and $F(d, q) = \sum_{j=0}^{N-1} c^4 q^2 e^{-2q(d + jw)}/16$ filters polarization fluctuations at different wavevectors. 
Crucially, $F(d, q)$ is sharply peaked at $q=1/d$ and so the qubit will only be affected by fluctuations in the polarization around this wavevector.
By combining Eqs.~\eqref{eq-1/T1_EEresonant}~and~\eqref{eq-EE_from_C}, we find that:
\beq\label{eq-1/T1Full}
\frac{1}{T_1} = \frac{d_{\perp}^2 \mu_0^2}{2} \coth\left( \frac{\beta \omega_q}{2} \right)  \int\frac{d^2 \q}{(2\pi)^2}F(d,q) \mathcal{C}(\omega_q, \q)~~~~
\eeq
Therefore, by tuning the qubit's frequency splitting $\omega_q$ and the qubit-sample distance $d$, one can effectively reconstruct $\mathcal{C}(\omega, \q)$ \cite{ftShubhayu}, giving one access to the dielectric properties of a proximate material.

A few remarks are in order.
First, we note that existing qubit sensing setups have demonstrated the capability to tune a probe qubit's frequency between $10\ \text{MHz} - 10\ \text{GHz}$,  qubit-sample distances down to $\sim 10$\ \text{nm}, and temperatures between $1-600~\text{K}$~\cite{Myers2017,Breeze2018,Maletinsky2012}.
The parameter regimes accessible by qubit sensors and other (near-)equilibrium probes of polar and dielectric materials are depicted in Fig.~\ref{fig:Fig_1}(b)~\cite{SM, SHGReview_McCullian2020, XRLD_Polisetty_2012, PFM_Review_Gruverman2019, XPCS_Grbel2008, TEM_Winkler2012, DielectricSpectroscopy_Grigas2019, Parsonnet2020, pump_probe_ft}, highlighting the complementary nature of our probe to existing experimental techniques.
Second, note that the frequency scales accessible to qubit sensors are small relative to the excitation energy scales of typical materials ($\sim \textrm{eV}$).
Thus, they will be sensitive to gapless or weakly gapped polar excitations.

The ability to probe such excitations naturally enables qubit sensors to address questions about polar and dielectric materials relevant to both fundamental and applied science.
We examine in detail three such questions.


\textbf{Detecting Polar Collective Modes.} We start discussing how qubit sensors can detect collective modes in neutral polar fluids.
While the existence of ``plasmon'' collective modes, arising from long-range Coulomb interactions between electrons in metals, is well established~\cite{ssbook}, the conclusive observation of their dipolar analogues (``dipolarons'') has remained an outstanding challenge~\cite{AlderPollock,Ascarelli76,Chandra90}. 
Dipolarons in a 2D dipolar fluid with density, $n_d$, molecular mass, $m$, and dipole moment, $\boldsymbol{\mu}$, are predicted to be gapless~\cite{SM,AlderPollock} with an unusual dispersion $\omega_d^2(\q) = v^2 q^2 + 2\pi n_d q (\q \cdot \boldsymbol{\mu})^2/m$, which is anisotropic due to the directional dependence of the dipolar interaction.
With this dispersion, we can derive the polarization correlation function $\chi(\omega, \q) \sim \omega_d^2(\q)/[\omega^2 - \omega^2_d(\q)]$, and hence predict the frequency and distance scaling of $1/T_1$ for a nearby qubit \cite{SM, ft3}.
%
\beq \label{eq-1/T1Gauss}
\frac{1}{T_1}\sim \coth\left(\frac{\beta \omega_q}{2}\right) \times  \left[ e^{-2q_{res}d}q_{res}^2 \right]  \Theta( \omega_q - \omega_0)
\eeq
where $q_\text{res}$ satisfies the resonance criterion $\omega(q_{\text{res}}) = \omega_q$.
Thus, for gapless dipolarons, the crossover from a linear to $q^{3/2}$ dispersion with increasing $q$ manifests in a corresponding crossover in the frequency scaling of $1/T_1$ from $\omega_q e^{-2 \omega_q d/v}$ to $\sim \omega_q^{1/3} e^{- 2(\omega_q)^{2/3}d}$  (for $T \gg \omega_q$), and can serve as a smoking gun signature of dipolarons.

\textbf{Probing Criticality with Spectating Modes.} The ability to probe low-energy polar excitations further enables qubit sensors to characterize phase transitions in polar and dielectric materials. 
Here, we make predictions for the behavior of $1/T_1$ across continuous instances of such transitions.
Our predictions are motivated two-fold.
First, there are many outstanding fundamental questions that remain surrounding central peak phenomena \cite{bruce1980structural,hushur2007}, quantum critical behavior \cite{Pavlova2009,rowley2014ferroelectric}, and thin-film polar instabilities~\cite{Blinov2000,cheema2020enhanced}.
Moreover, several recently discovered 2D ferroelectric and multiferroic compounds cannot be probed by standard bulk techniques \cite{wang2023towards}, motivating the development of nanoscale electric sensors. 

\begin{figure}
    \centering
    \includegraphics[width = 247pt]{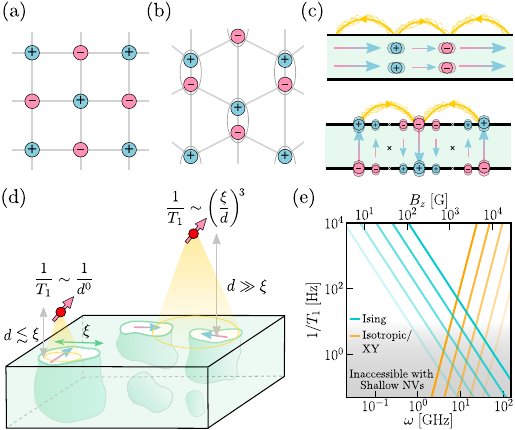}
    \caption{\textbf{Applications and Feasibility of Qubit Relaxometry.} We explore different applications of qubit relaxometry and assess the feasibility of exploring them in contemporary experiments. 
    For example, such sensors can characterize transitions between different polar phases, e.g. between paraelectrics (a) and ferroelectrics (b).
    These transitions are probed by coupling to ``spectating modes'' that appear near transitions, shown schematically for isotropic/XY anisotropic (c, top) and Ising anisotropic (c, bottom) materials.
    (d) Moreover, the physics of disordered ``relaxor'' ferroelectrics---often characterized by the formation of randomly-oriented polar domains---can also be probed; the characteristic size $\xi$ and dynamics of these domains are ascertained by the distance and frequency dependence of $1/T_1$.
    (e) To assess experimental feasibility, we make predictions for the relaxation rate arising from phase transitions in Ising and XY anisotropic/isotropic materials.
    Estimates are shown for values of $t \Omega_0^2$ (defined in the main text) ranging from $10^{-2}$\,--\,$10^{-4}$ $\mathring{\text{A}}$ eV$^2$ (light to dark) and material parameters defined near Eq.~\eqref{eq:t1xy}.
    %
    }
    \label{fig:Fig_2}
\end{figure}

Let us recall that the paraelectric-to-ferroelectric (PE/FE) transition is a structural phase transition accompanied by inversion-symmetry breaking and a spontaneously generated polarization density. 
It can be visualized by considering an ionic crystal with alternating charges $\pm Q$ shown in both the PE and FE phase in Fig.~\ref{fig:Fig_2}(a). 
In isotropic 3D bulk materials, this transition is driven by the condensation of the transverse optical (TO) phonon mode---relative displacements $\mathbf{u}$ between the $\pm Q$ charges depicted---which occurs either due to quantum or thermal fluctuations: depending on whether the transition is driven by temperature (a ``thermal phase transition'') or a separate tuning parameter $\lambda$ (e.g., strain) at $T=0$ (a ``quantum phase transition'')~\cite{ssbook,SondhiRMP}.
%
Unfortunately, owing to the transverse nature of such modes (i.e., $\mathbf{q} \cdot \mathbf{u} \propto \mathbf{q} \cdot \mathbf{P} = 0$), they generate no (bound) electric charge density and hence do not directly generate electric fields. 
As such, naively, such modes cannot be probed by qubit sensors!

However, the interfaces between the sample, probe, and substrate are known to induce \cite{vinogradov1992,jardin2002} additional ``spectating'' polar modes, which are neither strictly transverse nor longitudinal.
Crucially, these modes have the 
property that they (1) become gapless at the PE/FE transition (hence are sensitive to the criticality) and (2) directly generate stray electric fields outside the sample.
Using such modes, we now demonstrate that it is possible to extract certain exponents of the PE/FE transition.
%

To describe these modes and the associated fluctuating electric field, we consider a thin slab of a polar or dielectric material with thickness $t$, sandwiched between a probe material and a substrate with dielectric constants $\varepsilon_p$ and $\varepsilon_s$ respectively. The dielectric constant of the material is assumed to be dominated by a soft mode $\varepsilon \propto (\omega^2-\omega_{TO}^2)^{-1}$.
In such a setting, one can use Maxwell's equations \cite{SM} to demonstrate that two distinct types of spectating modes arise depending on the anisotropies of the samples in-plane and out-of-plane dielectric functions $\varepsilon_{\parallel}$ and $\varepsilon_{\perp}$. 
In what follows, we primarily discuss modes that arise in isotropic ($\varepsilon_{\parallel} = \varepsilon_{\perp}$) or easy-plane materials ($\varepsilon_{\parallel} > \varepsilon_{\perp}$), before returning to the Ising anisotropic case ($\varepsilon_{\parallel} < \varepsilon_{\perp}$) where the nature of the PE/FE transition is dramatically altered by surface effects.

In the isotropic/easy-plane anisotropic case, the surface leads to spectating modes [visualized in Fig.~\ref{fig:Fig_2}(c, top)] that 
have a dispersion: 
\begin{equation}
    \omega^2(\mathbf{q}) = \omega_T^2(\mathbf{q}) + a |\mathbf{q}| \qquad a = \frac{\varepsilon_{\infty}\omega_L^2 t}{\varepsilon_s + \varepsilon_p}
    \label{eq:IsoDisp}
\end{equation}
where $\omega_{T/L}(q)$ are the bulk TO/LO phonon dispersions 
and $\varepsilon_{\infty}$ is the material's bare dielectric constant.
%
The $q$-linear portion of $\omega^2(q)$ arises because the longitudinal component of these modes generates electric charge density implying their dispersion incurs a correction due to the Coulomb interaction.
Moreover, its gap tracks the gap of the TO phonon mode, analogous to how the $q = 0$ LO phonon frequency is equal to TO phonon frequency at $q=0$ in the 2D limit \cite{sohier2017}. 

The physics of the Ising case (polarization develops perpendicular to the film) is quite different. The $q=0$ mode is now purely longitudinal; therefore, increasing $q$ lowers the energy of the mode, implying that instability occurs at a finite $q$ \cite{SM}, i.e., $\omega^2(\mathbf{q}) \approx \omega_0^2 +c_s^2 (q-q_0)^2$ for sufficiently small $q_0$ such that anisotropies can be neglected. In the ordered state, this implies formation of domains of size $\sim1/q_0$ \cite{ashcroft1976solid}. This alters the universality class of the transition. While this is known theoretically in the context of dipolar Ising magnets \cite{saratz2016,cannas2018}, the finite-momentum resolution of qubit sensors could enable the experimental detection of this universality class.

Let us discuss the contribution of the critical spectating modes to $1/T_1$. For purely harmonic phonons, a nonzero contribution arises once the mode frequency ($\omega_{T}(0)$ or $\omega_0$) becomes smaller than the frequency splitting of the qubit. For XY/isotropic case, the depolarization rate scales as $1/T_1 \propto \coth(\beta \omega_q/2) q_{\rm res}^3 e^{-2 q_{\rm res} d}$, where $q_{\rm res}$ corresponds to the momentum where the mode-dispersion is resonant with the qubit frequency, i.e., $\omega(q_{\rm res}) = \omega_q$.
Therefore, one may extract $q_{\rm res}$ and hence the gap of the critical TO mode $\omega_{0,T} = \sqrt{\omega_q^2 - a q_{\rm res}}$ from the $1/T_1$, thereby tracking the approach to criticality. Below we provide $1/T_1$ estimates based on this approach (for results see Fig. \ref{fig:Fig_2} (e)).

On approaching the true criticality, however, one expects the nonlinearities to qualitatively alter the behavior of polar correlations. In the End Matter we show that the dynamical correlations of the spectator mode will also reflect these critical fluctuations, allowing to determine a number of critical scaling parameters of both thermal and quantum polar phase transitions with relaxometry.

\textbf{Probing Dynamics of Relaxor Ferroelectrics.} 
%
%
Disorder and inhomogeneity are also known to affect dynamics of polar materials, most vivid in the case of relaxor ferroelectrics --- dielectric materials with anomalously large internal polar fluctuations resembling ``disorder broadened'' critical correlations of phase transitions 
\cite{cross1987relaxor,Smolensky1984,cohen2006relaxors,cowley2011relaxing, verri2012theory}. A full microscopic description of relaxors remains at present elusive~\cite{Varma1,Varma2, verri2012theory}; below we demonstrate that qubit sensors may provide new critical information about their behavior.

To discuss the main expected features, we rely on the picture of a relaxor ferroelectric as a state with polar nano-regions---nanoscale domains with non-zero spontaneous polarization pinned by disorder \cite{JeongPNR,pasciak2012polar,eremenko2019local}.
This physics is captured by a minimal classical model of dipoles each with dipole moment $\p$ with volume density $n$, embedded in a material with dielectric constant $\varepsilon$.
For isotropic fluctuations, the dynamic correlation function between the dipoles can be written as $\langle p_i(\r, t) p_j(\r^\prime, 0) \rangle = \frac{|\p|^2}{3} \delta_{ij} C(\r, \r^\prime,t)$. 
Inside polar nano-regions of size $\xi$, we expect locally correlated dipoles, each undergoing slow collective relaxational dynamics, which we model via simple Debye relaxation at a timescale $\tau$, i.e., $C(\r, \r^\prime, t) = e^{- |\r - \r^\prime|/\xi} e^{-t/\tau} \theta(t)$.
Subsequently, we evaluate the distance-scaling behavior of $T_1$ when the average distance between the dipoles is small compared to the qubit-sample distance, i.e., $n d^3 \gg 1$ 
\beq
\frac{1}{T_1} \propto \begin{cases}
   \left(\frac{\xi}{d}\right)^3 , ~ d \gg \xi \\
    d^0, ~ d \ll \xi
\end{cases}
\eeq
This scaling behavior has a simple intuitive explanation: at large distances $d \gg \xi$ the qubit senses electric field fluctuations from many dipolar domains with typical size $\xi^3$, while at small distances $d \lesssim \xi$ the qubit is sensitive to a single fluctuating domain [Fig.~\ref{fig:Fig_2}(d)], leading to a $T_1$ independent of distance~\cite{SM}. 
The crossover from $d^{-3}$ distance scaling of $T_1$ to a $d$-independent $T_1$ further enables us to determine the typical size of a coherently fluctuating dipolar domain or polar nano-region and the characteristic timescales, that can be also inhomogeneous in real materials \cite{takenaka2017slush}.
%
%
By studying $1/T_1$ of an isolated qubit as a function of in-plane coordinates at a fixed distance $d \lesssim \xi$
, a spatially resolved map of the static and dynamic domains could be obtained, evincing a microscopic description of relaxors.


\textbf{Experimental Realization and Feasibility.} Applications in mind, we now discuss a concrete realization of the qubit sensing setup and discuss its experimental feasibility.
In particular, we envision utilizing the NV center in diamond: a point defect consisting of a nitrogen substitution adjacent to a lattice vacancy defect.
The $^3A_2$ electronic spin manifold of the NV is modeled as a three-level system ($\ket{0}, \ket{+},$ and $\ket{-}$) and the degenerate $\ket{\pm}$ states are ideal for encoding the two-level qubit of our proposal~\cite{Maze2011}.
Crucially, these states can be initialized and manipulated optically and read-out through state-dependent fluorescence~\cite{Doherty2013}.
Moreover, as required, electric fields drive transitions between these states with dipole moment $d_{\perp} = 17\ \mathrm{ Hz}\cdot\mathrm{cm/V}$ and magnetic fields control their splitting with a magnetic moment of $\mu_z =  2.8 \text{ MHz}/\text{G}$~\cite{VanOort1990,Doherty2013}.
As a result, the splitting of these states can be controlled by local magnetic fields and the qubit-sample distance can be controlled by, for example, placing a nano-diamond with a single NV on a scanning probe tip~\cite{Ariyaratne2018,Finco2021,MagnonScatteringPlatform, Kolkowitz}, enabling probing $1/T_1$ as a function of both frequency and distance.

To assess the feasibility of this proposal, we express the relaxation rate of the NV as $T_1^{-1}= (T_1^\textrm{sig})^{-1} + (T_1^{\textrm{int}})^{-1}$. where $(T_1^\textrm{sig})^{-1}$ accounts for the signal from the sample and $(T_1^{\textrm{int}})^{-1}$ the intrinsic sources of relaxation from the diamond host; the latter establishes a limit on the sensitivity of our sensing protocol.
The magnitude of  $(T_1^{\textrm{int}})^{-1}$ has been reported in shallow NV samples ($\sim50~\textrm{nm}$ depth) as low as $10~\textrm{Hz}$ below $100~\textrm{K}$~\cite{Healey2020, ft2}.

We estimate \cite{SM} $1/T_1^{\text{int}}$ for ferroelectric materials near polar transitions, due to spectating modes as discussed above. For isotropic or in-plane polarization, we find that $\frac{1}{T_1^{\text{int}}}  \propto (\varepsilon_s+\varepsilon_p)^2$, as the density of states of the spectator mode is strongly affected by screening of the long-range Coulomb interaction. This result favors using thin films and high-$\varepsilon$ substrates. Using parameters of the polar modes typical for 2D materials \cite{sohier2017} and a high$-\varepsilon$ substrate, such as SrTiO$_3$ ($\varepsilon~ \cdot 10^4$ at low temperature $k_B T\sim 1$ THz \cite{yamada1969neutron}), rates of the order of a few Hz can be achieved already with $\omega_q\sim 10$ GHz (Fig. \ref{fig:Fig_2} e) (see End Matter and SM \cite{SM} for details).

We note that two recently discovered 2D ferroelectrics, SnTe \cite{chang2016discovery} and In$_2$Se$_3$ \cite{zheng2018room} show in-plane polarization, while their high-temperature paraelectric phases have rotational symmetry ($C_4$ and $C_3$, respectively) making the estimates above of potential relevance. 
From the above it follows that easy-plane ferroelectrics with low LO-TO splitting are best for NV sensing, which suggests potential applications to hyperferroelectrics \cite{garrity2014}, where this splitting is very small.

For the Ising case (polarization perpendicualr to the film/layer), as discussed above the universality class is altered due to the development of domains with size $\sim 1/q_0$. From the relaxometry perspective the resulting mode dispersion has an enhanced density of states which leads to a greatly enhanced signal with respect to the XY case (Fig. 2e). The abundance of material candidates, e.g. thin films of HfO$_2$ \cite{park2015ferroelectricity}, monolayers of CuInP$_2$S$_6$ \cite{liu2016room} and other Van der Waals materials \cite{wang2023towards} suggests a wide range of applicable materials.

\textbf{Conclusions.} In this work, we demonstrated that qubit sensors are a promising table-top tool for studying polar and dielectric materials down to the atomically thin limit
over a wide range of frequencies ($ \sim 10 \text{ MHz} - 10\text{ GHz}$) and temperatures ($1-600\ \text{K}$) with nanometer spatial resolution.
%
These capabilities make such sensors ideal for detection of low energy polar modes, enabling them to probe diverse physical phenomena.
%
%

We briefly comment on a few open directions involving qubit electrometry.
First, since previous work has demonstrated the sensing capabilities of impurity qubits at high pressures ($\sim \mathcal{O}(10)~\text{GPa}$), such qubits could naturally investigate the influence of stress and strain fields on polarization dynamics and could aid in characterizing strain-induced phase transitions~\cite{strain_PT_FE, strain_PT_FE_2, strain_PT_3}.
In addition, as illustrated in Eq.~\eqref{eq-1/T1Full}, our probe is more sensitive to surface physics at short sample-probe distances~\cite{SM} and could be used to resolve surface polarization dynamics, which can be distinct from the bulk~\cite{Binder}.
Furthermore, the nanoscale resolution of the qubit is ideally suited to probe unconventional ferroelectricity in moir\'e materials, which typically have superlattice lengthscales of tens of nanometers \cite{moire1,moire2}.
Finally, by using the electrical capabilities of qubit sensors, highlighted in this work, with the previously established magnetic capabilities, one may be able to probe the complex interplay between charge, polarization, and magnetization found in multiferroic materials \cite{RameshSkyrmions, flavian_2023,flavian_2024}.

\textbf{Acknowledgments.} 
We gratefully acknowledge discussions with Eugene Demler, Marcin Kalinowski, Francisco Machado, Thomas Mittiga, Joaquin Rodriguez-Nieva, Eric Peterson, Lokeshwar Prasad, and Chong Zu.
We would especially like to thank Eugene Demler, Francisco Machado, and Joaquin Rodriguez-Nieva for collaborations on related projects.
R.S. acknowledges support from the Barry M. Goldwater Scholarship,  UC Berkeley’s Summer Undergraduate Research Fellowship , and by the U.S. Department of Energy, Office of
Science, Office of Advanced Scientific Computing Research, Department of Energy Computational Science Graduate Fellowship under Award Number DESC0022158.
S.C. was supported by the ARO through the Anyon Bridge MURI program (grant number W911NF-17-1-0323)  and the  U.S. DOE, Office of Science, Office of Advanced Scientific Computing Research, under the Accelerated Research in Quantum Computing (ARQC) program.
E. P. acknowledges support from Intel Corporation under the FEINMAN Program. 
N.Y.Y. acknowledges support from the U.S. Department of Energy, Office of Basic Energy Sciences, Materials Sciences and Engineering Division, under Contract No.~DE-AC02-05-CH11231 within the High Coherence Multilayer Superconducting Structures for Large
Scale Qubit Integration and Photonic Transduction Program (QISLBNL). 
P. A. V. acknowledges support by the University of Connecticut OVPR Quantum CT seed grant.
L.W.M. acknowledges that this material is based upon work supported by the Air Force Office of Scientific Research under award number FA9550-24-1-0266. Any opinions, findings, and conclusions or recommendations expressed in this material are those of the author(s) and do not necessarily reflect the views of the United States Air Force.

\bibliographystyle{apsrev4-1} 
\bibliography{polar}

\begin{center}
{\bf\large End Matter}
\end{center}

\appendix*

\textbf{Noise Characterization of Critical Exponents.} 
The analysis presented in the main body of the paper indicates that qubit sensors can probe the soft-mode frequency by the $1/T_1$ enhancement. However, the discussion above did not take into account the critical physics around the transition, motivating a more careful analysis of $1/T_1$ around a critical point by using dynamical scaling theories for thermal and quantum transitions \cite{ssbook}. Here, we will focus on the case of films of isotropic or 
easy-plane materials.
In this case, the spectating mode is only one of many soft modes of the bulk and its effects on the transition can be neglected~\cite{SM}.

Near either a thermal or quantum critical point, the inverse of the dynamical polarization correlation function $C(\omega,\q)$ can be approximately decomposed into contributions from the critical TO mode fluctuations, given by the TO correlation function and from the long-range Coulomb interaction that contributes a non-analytic q-linear momentum dependence, i.e., $C^{-1}(\omega,\q)= G_{T}^{-1}(\omega,\q) -a |\q|$. While mean-field theory simply predicts a resonance at $\omega = \omega_{T,0}(q)$ for TO correlations ($G_{T,MF}^{-1}(\omega,\q) = \omega^2 - \omega_{T,0}^2(q)$), we expect that short-range interactions between critical TO modes will lead to a scaling form for their correlations upon approaching the transition \cite{ssbook}.

\textit{Predictions for Thermal Transitions---}For thermal phase transitions, this scaling function is primarily determined by the correlation length $\xi \propto |T - T_c|^{-\nu}$ and an energy scale $\omega_0 \sim |T - T_c|^{z \nu}$ that respectively diverge and vanish upon approaching the transition in a manner controlled by the correlation critical exponent $\nu$ and the dynamical critical exponent $z$~\cite{HH77}.
%
These two scales motivate studying $1/T_1$ in two regimes of the phase transition. 
The first $(\text{Regime I})$ is at the critical point $(T=T_c,\, \omega_0 \to 0,\,  \xi \to \infty)$, where the critical correlations are only sensitive to the ratio $\omega/(c_s q)^z$ and consequently $1/T_1$ will only depend on a specific ratio of the qubit frequency and the qubit-sample distance.
Meanwhile, the second $(\text{Regime II})$ is slightly away from the critical point but at large sample-qubit distance ($d\gg\xi$), corresponding to vanishing momenta $q \to 0$, where the correlations only depend on the ratio $\omega/\omega_0$.
Here, $1/T_1$ will be probed at large qubit-sample distances and only depends on the qubit frequency.
While the details of the scaling analysis are provided in the appendix, it yields\cite{omegalessthanT}:
\begin{equation}
    \frac{1}{T_1} \sim \begin{cases}\frac{2T}{d^{4 - \eta + z}} & \text{Regime I} \\ \frac{2 T}{ \omega_{0}^{(z + 2-\eta)/z} d^4} \bar{\Psi}\left(\frac{\omega_q}{\omega_{0}}  \right) & \text{Regime II}  \end{cases}
\end{equation}
where $\eta$ is the anomalous critical exponent that determines the spatial decay of $\eta$ controlling the anomalous decay of polarization correlations at criticality.
%
From this, a simple two-parameter scaling collapse enables one to extract all three critical exponents $(z, \nu, \eta)$ at the critical point \footnote{In particular, we remark that plotting $1/T_1$ in Regime I as a function of the distance enables extracting $\alpha = 4 - \eta + z$ from fitting. Similarly, in Regime II, we plot $|T - T_c|^{\nu (\alpha - 2)}/(2T \cdot T_1)$ as a function of $\omega_q/|T - T_c|^{\nu z}$ for different values of temperature.
We obtain $z$ and $\nu$ such that the curves for different temperatures collapse; with these $\eta = 4 + z - \alpha $, 
}. 

\textit{Predictions for Quantum Transitions---}The case of a quantum phase transition is similar.
Here, there are two distinct tuning knobs: the temperature $T$ and the distance from the critical point $|\lambda - \lambda_c|$, the latter of which controls the gap scale $\omega_0 \sim |\lambda - \lambda_c|^{z\nu}$. 
Considering similar regimes as earlier---Regime I being at the critical point and Regime II being slightly away but at large qubit-sample distances---we find, through a detailed scaling analysis, that:
\begin{equation}
    \frac{1}{T_1} \sim \begin{cases} \frac{1}{d^{4 - \eta}} \Phi_1\left(dT^{1/z}  \right) & \text{Regime I} \\ T^{(-2 + \eta)/z} \Phi_2\left( \frac{\omega_0}{T} \right) & \text{Regime II}  \end{cases}
\end{equation}
for scaling functions $\Phi_1$ and $\Phi_2$.
Once again, a simple two-parameter scaling analysis enables extracting all three critical exponents $(z, \nu, \eta)$ at the critical point.

\textbf{Estimates of $1/T_1$.}
In this section, we provide further technical details for our estimates of the qubit lifetime $1/T_1$ due to electrical fluctuations from proximate sample. 
Assuming that qubit is sufficiently close to the surface $q_{res}d \ll 1$, we have (see SM~\cite{SM} for a detailed derivation):
\begin{equation}
\frac{1}{T_1^{\text{int}}} 
= \frac{3\pi d_\perp^2 (\varepsilon_s+\varepsilon_p)^2}{t^3} \coth(\omega_q/2 T) \frac{\omega_q^6}{\Omega_0^6},
\label{eq:t1xy}
\end{equation}
where $\Omega_0$ is defined via the bulk dielectric constant $\varepsilon(\omega\to \omega_{TO})\approx \frac{\Omega_0^2}{\omega^2_{TO}-\omega^2}$. It is clear, that this result favors using thin films and high-$\varepsilon$ substrates  which enhance the density of states for the spectating mode. For 2d monolayer materials the role of $t\Omega_0^2$ is played by a single constant $\mathcal{S}$ \cite{sohier2017}, which takes values of $10^{-2}$\,--\,$10^{-4}$ $\mathring{\text{A}}$ eV$^2$, leading to $1/T_1 [\mathrm{Hz}] \gtrsim 10^{-2}\text{\,--\,}~10^{4} (\varepsilon_s+\varepsilon_p)^2 T \omega^5$ [THz]. Thus, rates of the order of a few Hz can be achieved already with $\omega_q\sim 10$ GHz (Fig. \ref{fig:Fig_2} e) using a high$-\varepsilon$ substrate, such as SrTiO$_3$ ($\varepsilon~ \cdot 10^4$ at low temperature $k_B T\sim 1$ THz \cite{yamada1969neutron}).

For the Ising case, as discussed above the universality class is altered. The instability occurs at a finite $q=q_0$ (due to gap vanishing at a finite $q$ value: $\omega^2(\mathbf{q}) \approx \omega_0^2 +c_s^2 (q-q_0)^2$) leading to the development of domains with size $\sim 1/q_0$. From the relaxometry perspective the resulting mode dispersion has an enhanced density of states which leads to

\begin{equation}
    \frac{1}{T_1} = \frac{3 d_\perp^2 \pi^3 \Omega_{0,\perp}^2}{\varepsilon_\parallel^2 t^3 c_sq_0 \omega}
    \coth\left( \frac{\beta \omega}{2} \right)  e^{-2q_0d}
\end{equation}
for thick films $q_0t\gg 1$; for 2D films and materials the result is multiplied by $\sim(q_0t)^4$.
At low frequencies, the signal from uniaxial ferroelectrics should be thus greatly enhanced with respect to the XY case (we note that the actual divergence in the $\omega\to 0$ limit would be cut off by anisotropies not included here). 


\newpage
$\ $
\newpage

\pagebreak
\onecolumngrid
\appendix

\section{Organization of Supplemental Materials}

The supplemental materials below are organized as follows. 
In Section~\ref{sec:derivation}, we provide a detailed derivation of the relaxation rate $1/T_1$ of a qubit sensor proximate to a polar and dielectric material, relating this rate to the polarization correlations in the material.
Sections~\ref{sec:spec},~\ref{sec:relaxspec},~and~\ref{sec:phasetransition} provide detailed calculations underlying our analysis of probing polar phase transitions.
In  particular, in Sections~\ref{sec:spec}~and~\ref{sec:relaxspec} we derive the existence of spectating soft modes in polar materials and derive the relaxation rate associated with these spectating modes, respectively.
The latter is further used in the feasibility analysis presented in the main text.
Section~\ref{sec:phasetransition} provides details on the scaling theory used to describe how qubit sensors can extract the critical exponents of a phase transition.

Sections~\ref{sec:dipolaron}~and~\ref{sec:relaxor} are devoted to the other two applications presented in our work.
In particular, the former discusses a derivation of the dipolaron mode spectrum and the contributions of these dipolarons to the polarization correlations appearing with the formula for $1/T_1$.
The latter discusses our derivation of the noise from the toy model for relaxor ferroelectrics presented in the main text.


\section{Derivation of Qubit Relaxation Rate}~\label{sec:derivation}

In this section, we systematically derive the relaxation rate of an impurity qubit sensor proximate to a polar or dielectric material in both a general setting and specific settings of interest. We start by deriving a relationship between the relaxation rate of the qubit and electrical noise at its location. Subsequently, we utilize Maxwell's equations to connect this electrical noise to polarization correlations in the nearby material. After this, we express the form of $1/T_1$ in a number of settings of interest. Finally, we investigate the influence of magnetic noise for the specific case of the NV qubit.

\subsection{Impurity Qubit Response to Electrical Noise}
Recall that in the main text, we defined the impurity qubit's coupling to electric and magnetic fields as (setting $\hbar = 1$ henceforth): 
\beq
H =  H_0 + H_{q-EM} =  \frac{\omega_q}{2} \sigma^z +  \hat{\mathbf{d}} \cdot \mathbf{E} + \hat{\boldsymbol{\mu}} \cdot \mathbf{B}, \quad  \quad \hat{\mathbf{d}} = d_{\perp} (\sigma^x, \sigma^y, 0) \text{ and } \hat{\boldsymbol{\mu}} = \mu_z (0, 0, \sigma^z)
\eeq
where $d_{\perp}$ was the electrical dipole moment, $\mu_z$ is the magnetic moment, and we assume that the quantization axis of the qubit is aligned with the physical $z$-axis of the system, defined as the axis normal to the plane of a proximate material. 
We assume that our sample is in thermal equilibrium at inverse temperature $\beta$ with density matrix $\rho = \frac{1}{Z} \sum_n e^{- \beta \varepsilon_n} \ket{n} \bra{n}$, where $\ket{n}$ is an eigenstate of the sample at energy $\varepsilon_n$ and $Z = \sum_n e^{-\beta \varepsilon_n}$ is the partition function.
Now, we use Fermi's golden rule to compute the transition rate between $\ket{1}$ to $\ket{0}$ (we set $\hbar = 1$ and $k_B = 1$ henceforth): 
\begin{align}\label{eq-FGR}
\Gamma_{em} &= 2\pi d_{\perp}^2 \sum_{nm} \frac{e^{-\beta \varepsilon_n}}{Z} |\bra{m, 0} E^x \sigma^x + E^y \sigma^y \ket{n, 1}|^2 \delta ( \omega_q - (\varepsilon_m - \varepsilon_n)) \\ 
 &= 2 \pi d_{\perp}^2 \sum_{nm} \frac{e^{-\beta \varepsilon_n}}{Z} |\bra{m, 0} E^x + iE^y \ket{n, 0}|^2 \delta ( \omega_q - (\varepsilon_m - \varepsilon_n)) \\
 &= 2 \pi d_{\perp}^2 \sum_{nm} \frac{e^{-\beta \varepsilon_n}}{Z} E^+_{mn} E^-_{nm} \delta ( \omega_q - (\varepsilon_m - \varepsilon_n))
\end{align}
where the argument of the delta enforces energy conservation, i.e, the amount of energy lost by the qubit ($\omega_q = E_1 - E_{0}$) equals the amount of energy gained by the sample $(\varepsilon_m - \varepsilon_n)$. Similarly, we have that: 
\beq
\Gamma_{abs} =  2\pi d_{\perp}^2 \sum_{nm} \frac{e^{-\beta \varepsilon_n}}{Z} E^-_{mn} E^+_{nm} \delta ( \omega_q + (\varepsilon_m - \varepsilon_n))
\eeq
where $E^{\pm} = E_x \pm i E_y$. Thus, we can write $1/T_1 = 1/2(\Gamma_{abs} + \Gamma_{em})$ \cite{Agarwal2017}. Now, to relate this quantity to the electric field fluctuations, note that the noise tensor is given by:
\begin{equation}
\mathcal{N}_{ij}(\omega) = \frac{1}{2} \int_{-\infty}^{\infty} \text{d}t \langle \{ E^{i}(t), E^{j}(0) \} \rangle e^{i \omega t} = \pi \sum_{nm} \frac{e^{-\beta \varepsilon_n}}{Z} \left[E_{nm}^i E_{mn}^j \delta(\omega - (\varepsilon_{m} - \varepsilon_n)) + E_{nm}^j E_{mn}^i \delta(\omega - (\varepsilon_{n} - \varepsilon_m))\right]~~~~~~~~~~~~
\end{equation}
Thus, it follows that: 
\begin{equation}\label{eq-T1NoiseTen}
\frac{1}{T_1} = d_{\perp}^2\mathcal{N}_{-+}(\omega_q)
\end{equation}
Subsequently, by the fluctuation-dissipation theorem
\begin{equation}\label{eq-FlucDissap}
\mathcal{N}_{ij}(\omega) = \frac{1}{2} \coth\left( \frac{\beta \omega}{2} \right)\mathcal{S}_{ij}(\omega) \text{ where } \mathcal{S}_{ij}(\omega) = \int_{-\infty}^{\infty} \text{d}t \langle [E^{i}(t), E^{j}(0)] \rangle e^{i \omega t}
\end{equation}
Moreover, we can relate $\mathcal{S}_{ij}(\omega)$ in terms of the retarded correlators of the electric field, which are more convenient to calculate
\begin{equation}\label{eq-SpectralDensityRetCorr}
\mathcal{S}_{ij}(\omega) = 2~\text{Im}\left[C^R_{E^i E^j}(\omega)\right] \text{ where } C_{E^i E^j}^R (\omega) = i \int_{-\infty}^{\infty} \text{d}t \, \Theta(t) \langle [E^i(t), E^j(0)] \rangle e^{i \omega t} 
\end{equation}

\subsection{Propagation of Maxwell's Equations}

To determine the electrical noise arising from dipolar fluctuations, we propagate these fluctuations using Maxwell's equations in Lorentz gauge: 
\beq
\partial^2 A^{\mu}(\r, t) = \mu_0 J^{\mu} (\r, t) = \mu_0 \begin{pmatrix} -c \nabla \cdot \mathbf{P}(\r, t) + c \sigma(\r, t) \\ \partial_t \P(\r, t) \end{pmatrix}
\eeq
where $\partial^2 = -\partial_t^2/c^2 + \nabla^2$, $\mathbf{P}(\r, t) = \mathbf{P}(\r, t) 1_{[-w, 0]}(z)$ (where $1_{[-w, 0]}$ is $1$ for $z \in [-w, 0]$ and $0$ otherwise), $\sigma(\r , t) = P_z(\r, t) \delta(z) - P_z(\r, t) \delta(z + w)$ is the surface charge density, and $w$ is the width of the sample. We can solve these equations by introducing a kernel $G_i^\mu(\r, \r', t - t')$: 
\beq
A^{\mu}(\r, t) = \mu_0 \int_{-\infty}^{\infty} dt' d^3 \r' G_i^{\mu} (\r, \r', t - t') P_i(\r', t')
\eeq
where $i$ labels $x, y, z$ and we are implicitly summing over repeated indices. We define $G_i^{\mu}$ to satisfy the equation: 
\beq
\partial^2 G_i^{\mu}(\rho - \rho', z, z', t-t') = \begin{pmatrix} c\delta(t - t')  \partial_i[\delta^{(3)} (\r - \r')] + c \delta_{i, z} \delta^{(3)} (\r - \r') [\delta(z) - \delta(z + w)] \\ -\delta^{(3)} (\r - \r') \partial_t[\delta(t - t')]  \hat{e}_i\end{pmatrix}
\eeq
where $\rho = (x, y)$ is the coordinates of the material in-plane. Now, to solve for $G_i^{\mu}$, we can express $A^{\mu}, G_i^{\mu},$ and our polarization $P_i$ in terms of their in-plane Fourier modes yielding: 
\beq
A^{\mu}(\r, t) = \frac{1}{\sqrt{L^2}} \sum_{\q} \int \frac{d\omega}{2\pi} A^{\mu}(z, \q, \omega) e^{i(\q \cdot \rho - \omega t)}
\eeq
\beq
G^{\mu}_i(\rho, z, z', t) = \frac{1}{L^2} \sum_{\q} \int \frac{d\omega}{2\pi} G_i^{\mu}(z,z', \q, \omega) e^{i(\q \cdot \rho - \omega t)}
\eeq
\beq
P_i(\r, t) = \frac{1}{\sqrt{L^2}} \sum_{\q} \int \frac{d\omega}{2\pi} P_i(z, \q, \omega) e^{i(\q \cdot \rho - \omega t)}
\eeq
where we assumed a sample with transverse dimensions $L \times L$ for simplicity.
When we plug this back into the equations of motion for $G_i^{\mu}$, we get: 
\beq 
(-\lambda^2 + \partial_z^2) G_i^\mu (z, z', \q, \omega) = \begin{pmatrix} ic q_i (\delta_{i,x} + \delta_{i,y}) \delta(z - z') + c\delta_{i,z} \partial_z [\delta(z - z')] + c \delta_{i, z} \delta(z - z') [\delta(z) - \delta(z + w)] \\ i \omega \delta(z - z') \hat{e}_i\end{pmatrix}~~~~~~~~
\eeq
where $\lambda^2 = (q^2 - \omega^2/c^2)$. To solve this, we Fourier transform the $z$ coordinate as
\beq
G_i^{\mu}(\alpha, z', \q, \omega) = \int dz\, e^{-i \alpha z} G_i^{\mu}(z, z', \q, \omega)
\eeq
and so we get the following:
\beq
G_i^{\mu}(\alpha,z', \q, \omega) = -\frac{1}{\lambda^2 + \alpha^2} \begin{pmatrix} i c q_i (\delta_{i,x} + \delta_{i,y})e^{-i \alpha z'} + i c \delta_{i,z} \alpha e^{-i \alpha z'} + c\delta_{i,z} \int dz e^{-i \alpha z} \delta(z - z') \left[ \delta(z) - \delta(z + w) \right] \\ i \omega e^{-i \alpha z'} \hat{e}_i \end{pmatrix}~~~~~~~~
\eeq
Now, we Fourier transform back to get a useable expression for $G$. We do this one component at a time:
\beq
G_i^0 = - \frac{i c q_i (\delta_{i,x} + \delta_{i,y})}{2\lambda} e^{- \lambda |z -z'|}  - \frac{c \delta_{i,z}}{2} \text{sgn}(z - z') e^{-\lambda |z - z'|}- c \delta_{i, z} \int \frac{d\alpha d\tilde{z}}{2\pi}   \frac{e^{i \alpha (z - \tilde{z})}\delta(\tilde{z} - z') [\delta(\tilde{z}) - \delta(\tilde{z} + w)] }{\lambda^2 + \alpha^2} ~~~~~~~~\\
=- \frac{i c q_i (\delta_{i,x} + \delta_{i,y})}{2\lambda} e^{- \lambda |z -z'|}  - \frac{c \delta_{i,z}}{2} \text{sgn}(z - z') e^{-\lambda |z - z'|}- c \delta_{i, z} \int d\tilde{z} \frac{e^{-\lambda |z - \tilde{z}|}}{2\lambda} \delta(\tilde{z} - z') \left[\delta(\tilde{z}) - \delta(\tilde{z} + w) \right]~~~~~~~~~\\
=- \frac{i c q_i (\delta_{i,x} + \delta_{i,y})}{2\lambda} e^{- \lambda |z -z'|}  - \frac{c \delta_{i,z}}{2} \text{sgn}(z - z') e^{-\lambda |z - z'|}- \frac{c \delta_{i, z} e^{-\lambda |z - z'|}}{2 \lambda} [\delta(z') - \delta(z' + w)]~~~~~~~~
\eeq
Also, we have that:
\beq
G_i^{j} = -i \omega \frac{e^{-\lambda |z - z'|}}{2 \lambda} \delta_i^j
\eeq
where $i, j \in \{x, y, z\}$. Now, we can decompose our Green's function as: 
\beq
G_i^{\mu}(z, z', \q, \omega) = \mathcal{G}_i^{\mu}(z - z', \q, \omega) + g_i^{\mu}(z, z', \q, \omega)
\eeq
where
\beq
\mathcal{G}_i^{\mu}(z - z', \q, \omega) = \begin{pmatrix} - \frac{i c q_i (\delta_{i,x} + \delta_{i,y})}{2\lambda} e^{- \lambda |z -z'|}  - \frac{c \delta_{i,z}}{2} \text{sgn}(z - z') e^{-\lambda |z - z'|} \\ -i \omega \frac{e^{-\lambda |z - z'|}}{2 \lambda} \hat{e}_i \end{pmatrix} \label{eq-referenceG} \\  
g_i^{\mu} = \begin{pmatrix} - \frac{c \delta_{i, z} e^{-\lambda |z - z'|}}{2 \lambda} [\delta(z') - \delta(z' + w)] \\ 0 \end{pmatrix} \label{eq-referenceg}
\eeq
indicate bulk and surface terms respectively. Green's function in hand, we can relate the polarization back to the vector potential as: 
\begin{align}
A^{\mu}(\r, t) &= \frac{\mu_0}{L} \sum_\q \int \frac{dz' d\omega}{2\pi} G_i^{\mu}(z, z', \q, \omega) P_i (z', \q, \omega) e^{i (\q \cdot \rho - \omega t )}\\&= \frac{1}{L} \sum_\q \int \frac{dz' d\omega}{2\pi}  \mathcal{G}_i^{\mu} (|z-z'|, \q, \omega) P_i (z, \q, \omega) e^{i (\q \cdot \rho - \omega t )} + \frac{1}{L} \sum_\q \int \frac{dz' d\omega}{2\pi} g_i^{\mu}(z, z', \q, \omega) P_i (z', \q, \omega) e^{i (\q \cdot \rho - \omega t )} 
\end{align}
Therefore, we can compute the electric field as:
\beq
\bo{E}(\r, t) = - c\mathbf{\nabla} A^0(\r, t) - \partial_t \bo{A}(r, t) =  \frac{\mu_0}{L} \sum_q \int \frac{d \omega dz'}{2 \pi} \bo{H}_i(z,z', \q, \omega) P_i(z', \q, \omega) e^{i (\q \cdot \rho - \omega t)} 
\eeq
where 
\beq
\bo{H}_i = \begin{pmatrix} -i cq_x G_i^0 + i \omega G_i^x \\ -i cq_y G_i^0 + i \omega G_i^y \\ -c\partial_z G_i^0 + i \omega G_i^z \end{pmatrix} = \boldsymbol{\mathcal{H}}_i + \mathbf{h}_i
\eeq
For reference, the explicit form of the bulk kernels is:
\beq
\boldsymbol{\mathcal{H}}_x = -\frac{e^{-\lambda|z-z'|}}{2\lambda}\begin{pmatrix} c^2q_x^2 + \omega^2 \\ c^2q_x q_y \\ ic^2q_x \lambda \cdot \text{sign}(z-z')  \end{pmatrix} \quad \boldsymbol{\mathcal{H}}_y =-\frac{e^{-\lambda|z-z'|}}{2\lambda}\begin{pmatrix} c^2q_x q_y \\ c^2q_y^2 + \omega^2 \\ i c^2q_y \lambda \cdot \text{sign}(z-z') \end{pmatrix} 
\eeq
\beq
\boldsymbol{\mathcal{H}}_z = - \frac{e^{-\lambda |z - z'|}}{2}\begin{pmatrix}-ic^2q_x \text{sign}(z-z') \\-ic^2q_y \text{sign}(z-z') \\  - c^2\delta(z-z') + \frac{c^2\lambda^2\text{sign}(z-z')+ \omega^2}{\lambda}  \end{pmatrix}
\eeq
and the explicit form of the surface kernels are:
\beq
\mathbf{h}_z =  \frac{c^2 e^{-\lambda|z - z'|}}{2 \lambda}\begin{pmatrix}iq_x \\ iq_y \\ -\partial_z\end{pmatrix}  [\delta(z') - \delta(z' + w)] = \frac{c^2 e^{- \lambda |z - z'|}}{2 \lambda} \begin{pmatrix}iq_x \\ iq_y \\ -\partial_z\end{pmatrix} \mathcal{S}(z')
\eeq
with $\bo{h}_{x,y} = 0$.

\subsection{Qubit Relaxation from Dipolar Fluctuations (Translation Invariant)}

Having propagated the in-sample polarization to the electric fields outside of the material, we now determine the relaxation rate of our probe qubit due to in-sample polarization fluctuations. To do so, we compute electrical noise at the location of the probe qubit due to these fluctuations. In particular, by Eq.~\ref{eq-T1NoiseTen}~and~\ref{eq-FlucDissap}, we need to compute: 
\beq
\langle [E_-(\r, t), E_+(\r, 0)] \rangle = \frac{\mu_0^2}{L^2} \sum_{\q_1, \q_2} \int \frac{d \omega_1 d\omega_2 dz_1' dz_2'}{(2 \pi)^2} H_i^-(z_1', \q_1, \omega_1) H_j^+(z_2', \q_2, \omega_2) \\  \times \langle[P_i(z_1',  \q_1, \omega_1), P_j(z', \q_2, \omega_2] \rangle e^{i(\q_1 \cdot \boldsymbol{\rho} - \omega_1 t)} e^{i\q_2 \cdot \rho}
\eeq
where $\r = (\rho, d) =  (0, 0, d)$ is the location of the qubit.
Assuming spacetime translation invariance, we have that
\beq
\langle[P_i(z_1', \q_1, \omega_1), P_j(z_2', \q_2, \omega_2)] \rangle = 2 \pi \delta(\omega_1 + \omega_2) \delta_{\q_1, -\q_2} \langle[P_i(z_1', \q, \omega), P_j(z_2', -\q, -\omega)] \rangle
\eeq
Thus, the electrical noise can be expressed as: 
\beq
\langle[E_-(\r, t), E_+(\r, 0)] \rangle = \frac{\mu_0^2}{L^2} \sum_{\q} \int \frac{d \omega dz_1' dz_2'}{2\pi} H_i^-(z_1', \q, \omega) H_j^-(z_2', -\q, -\omega) \langle [P_i(z_1', \q, \omega), P_j (z_2', -\q, -\omega)] \rangle e^{-i \omega t}~~~~~
\eeq
To proceed further, we simply need to contract the product of kernels with the polarization commutator:
\beq
H_i^-(1)  H_{j}^+(2) \langle [P_i(1), P_j(1)]\rangle = \left\langle \left[ \frac{1}{2} (H_+^-(1) P_-(1) + H_-^-(1) P_+(1)) + H_z^-(1) P_z(1), \right. \right.\\ \left.\left.\frac{1}{2} (H_+^+(2) P_-(2) + H_-^+(2) P_+(2)) + H_z^+(2) P_z(2)\right]\right\rangle \\ 
= \frac{1}{4} \left( H_+^-(1) H_-^+(2) \langle[P_-(1), P_+(2)]\rangle + H^-_-(1) H^+_+(2) \langle[P_+(1), P_-(2)]\rangle\right) + H^-_z(1) H^+_z(2) \langle [P_z(1), P_z(2)] \rangle \\
 + \frac{1}{4} \left(H_+^-(1) H_+^+(2) \langle[P_-(1), P_-(2)]\rangle + H_-^-(1) H_-^+(2) \langle[P_+(1), P_+(2)]\rangle \right) \\
 + \frac{1}{2} \left( H_+^-(1) H_z^+(2) \langle [P_-(1), P_z(2)] + H_-^-(1) H_z^+(2) \langle [P_+(1), P_z(2)] \rangle \right. \\  \left. +  H_z^-(1) H_-^+(2) \langle [P_z(1), P_+(2)] \rangle + H_z^- (1) H_+^+(2) \langle [P_z(1), P_-(2)] \rangle\right)
\eeq
where $H_i^-(1) = H_i^-(d, z_1, \q_1, \omega_1)$, $H_j^+(2) = H_j^+(d, z_2, \q_2, \omega'_2), P_i(1) = P_i(z_1, \q_1, \omega_1),$ and $P_j(2) = P_j(z_2, \q_2, \omega_2)$ with $\q_1 = -\q_2 = \q$ and $\omega_1 = -\omega_2 = \omega$ and also $H^{\pm}_{\pm} = H_x^{\pm} \pm i H^{\pm}_y$ and $P_{\pm} = P_x \pm i P_y$. Although the above expression looks daunting, the first line is the only line that appreciably contributes when either it is a good approximation that the polarization is conserved or when the sample is rotationally invariant. Now we compute the product of the kernels in the approximation that $\omega/c \ll q$ (i.e. the speed of light is much faster than any velocity scale in the material). We remark that: 
\begin{align}
H^+_+ = - \frac{e^{-\lambda |z - z'|}}{2\lambda} c^2(q_x + i q_y)^2 \quad 
H^-_- = - \frac{e^{-\lambda |z - z'|}}{2\lambda}& (q_x - i q_y)^2
 \quad H^+_- = H^-_+ = - \frac{e^{-\lambda |z - z'|}}{2 \lambda} \left(c^2 q^2 + 2\omega^2 \right) \\
H^{\pm}_z =  \frac{e^{-\lambda |z - z'|}}{2\lambda} ic^2 (q_x \pm i q_y)& \left[ \lambda  \cdot \text{sign}(z - z') + \mathcal{S}(z') \right]
\end{align}
Now, we can compute the products of these kernels. First, the polarization conserving kernels:
\begin{align}
H_-^-(1) H_+^+(2) &\approx \frac{1}{4} e^{-q|z - z_1'|}e^{-q|z - z_2'|} c^4 q^2  = F(z, q; z_1', z_2')\\ 
H_+^-(1) H_-^+(2) &\approx \frac{1}{4} e^{-q|z - z_1'|}e^{-q|z - z_2'|} c^4  q^2 = F(z, q; z_1', z_2')
\end{align}
and also
\begin{align}
H_z^-(1) H_z^+(2) &\approx \frac{1}{4} e^{-q|z - z_1'|}e^{-q|z - z_2'|} c^4 (q^2 + q (\mathcal{S}(z_1') + \mathcal{S}(z_2')) + \mathcal{S}(z_1') \mathcal{S}(z_2'))\\ &= F(z, q; z_1', z_2') \left[ 1 + \frac{1}{q} (\mathcal{S}(z_1') + \mathcal{S}(z_2')) +  \frac{1}{q^2} \mathcal{S}(z_1') \mathcal{S}(z_2') \right]
\end{align}
Now, the other terms:
\begin{align}
H_+^- (1) H_+^+(2) &\approx \frac{1}{4} e^{- q|z - z_1'|} e^{-q|z - z_2'|} c^4 q^2 e^{2i \phii} = F(z, q; z_1', z_2') e^{2i\phii} \\
H_-^- (1) H_-^+(2) &\approx \frac{1}{4} e^{- q|z - z_1'|} e^{-q|z - z_2'|} c^4 q^2 e^{-2i \phii} = F(z, q; z_1', z_2') e^{-2i\phii} \\
H_+^-(1) H_z^+(2) &\approx -\frac{1}{4} e^{- q|z - z_1'|} e^{- q|z - z_2'|} (-ie^{i \phii}) c^4 \left[q^2 + q \mathcal{S}(z_2') \right] = i e^{i \phii} F(z, q; z_1', z_2') \left[1 + \frac{1}{q} \mathcal{S}(z_2') \right] \\
H_-^-(1) H_z^+(2)  &\approx -\frac{1}{4} e^{- q|z - z_1'|} e^{- q|z - z_2'|} (-ie^{-i \phii}) c^4 \left[q^2 + q \mathcal{S}(z_2') \right] = i e^{-i \phii} F(z, q; z_1', z_2') \left[1 + \frac{1}{q} \mathcal{S}(z_2') \right]  \\
H_z^-(1) H_+^+(2) &\approx -\frac{1}{4} e^{- q|z - z_1'|} e^{- q|z - z_2'|} ie^{i \phii} c^4 \left[q^2 + q \mathcal{S}(z_1') \right] = -i e^{i \phii}F(z, q; z_1', z_2') \left[1 + \frac{1}{q} \mathcal{S}(z_1') \right]  \\
H_z^-(1) H_-^+(2)  &\approx -\frac{1}{4} e^{- q|z - z_1'|} e^{- q|z - z_2'|} ie^{-i \phii} c^4 \left[q^2 + q \mathcal{S}(z_1') \right]  = -i e^{-i \phii}F(z, q; z_1', z_2') \left[1 + \frac{1}{q} \mathcal{S}(z_1') \right]
\end{align}
where $\phii$ is the angle that $\q$ makes in the $xy$ plane. Now, we can write the full expression for the electrical noise:
\begin{align}
\langle [E_-(\r, t), E_+(\r, 0)] \rangle &= \frac{\mu_0^2}{L^2} \sum_{\q} \int \frac{d\omega dz_1' dz_2'}{2\pi} F(d, q; z_1', z_2') \times (\mathcal{C}_{bb} + \mathcal{C}_{bs} + \mathcal{C}_{ss}) e^{-i \omega t}
\end{align}
where $\mathcal{C}_{bb}$ are bulk-bulk correlations, $\mathcal{C}_{bs}$ are correlations between the bulk and the surface, $\mathcal{C}_{ss}$ are correlations between the two surfaces of the sample. Let us enumerate these one-by-one: 
\begin{align}
\mathcal{C}_{bb} &= \frac{1}{16} ( \langle [P_+(1), P_-(2)] \rangle + \langle [P_-(1), P_+(2)] \rangle + 4 \langle[P_z(1), P_z(2)] \rangle )\\
&+ \frac{1}{16} ( \langle [P_-(1), P_-(2) ] \rangle  e^{2 i\phii} + \langle [P_+(1), P_+(2)] \rangle e^{-2i \phii} ) \\
&+ \frac{1}{8} \left[ \left(\langle [P_-(1), P_z(2)] \rangle ie^{i \phii} +  \langle [P_+(1), P_z(2)] \rangle ie^{-i \phii} \right)- \left(\langle [P_z(1), P_-(1)] \rangle i e^{i \phii} +  \langle[P_z(1), P_+(2)] \rangle i e^{-i \phii}\right)\right]
\end{align}
where the parenthesis in the last term indicate complex conjugate pairs. Moreover, we have that:
\begin{align}
\mathcal{C}_{bs} & =  \frac{\mathcal{S}(z_1')}{8q} \left( 2\langle[P_z(1), P_z(2)] \rangle - i e^{i \phii} \langle [P_z(1), P_-(2)] \rangle - i e^{-i\phii} \langle [P_z(1), P_+(2)] \rangle  \right) \\ &+ \frac{\mathcal{S}(z_2')}{8q} \left( 2\langle[P_z(1), P_z(2)] \rangle + i e^{i \phii} \langle [P_-(1), P_z(2)] \rangle + i e^{-i\phii} \langle [P_+(1), P_z(2)] \rangle  \right)
\end{align}
Finally, we have the surface-surface correlations:
\begin{align}
\mathcal{C}_{ss} & =  \frac{1}{4q^2}\mathcal{S}(z_1')\mathcal{S}(z_2') \langle[P_z(1), P_z(2)] \rangle
\end{align}
So, we can write down our relaxation rate as:
\beq \label{eq-Final1/T1Expression}
\frac{1}{T_1} = \frac{1}{2}d_{\perp}^2 \coth\left( \frac{\beta \omega}{2} \right) \frac{\mu_0^2}{L^2} \sum_{\q} \int dz_1' dz_2' F(d, \q; z_1', z_2') \times \{ \mathcal{C}_{bb} + \mathcal{C}_{bs} + \mathcal{C}_{ss} \}
\eeq
Note that in the special case where the material is a stack of $N$ 2D layers each of width $w$, we can re-express our correlators as 
\beq
\langle [P_\alpha (z_1', \q, \omega), P_\beta(z_2', -\q, -\omega)] \rangle = \sum_{j = 0}^{N-1} \langle [P_\alpha (\q, \omega), P_\beta(-\q, -\omega)] \rangle \delta(z_1' -  jw) \delta(z_2' - j w)
\eeq
Consequently, the expression for $1/T_1$ can be re-written as: 
\beq \label{eq-Final1/T1Expression2D}
\frac{1}{T_1} = \frac{1}{2} d_{\perp}^2 \coth\left( \frac{\beta \omega_q}{2} \right) \frac{ \mu_0^2}{L^2} \sum_{\q} F(d,\q) \times  \{\mathcal{C}_{bb} + \mathcal{C}_{bs} + \mathcal{C}_{ss}\}
\eeq
where $F(d, \q) = \sum_{j=0}^{N-1} c^4 q^2 e^{-2q(d + jw)}$ and $\mathcal{C}_{bb}, \mathcal{C}_{bs},$ and $\mathcal{C}_{ss}$ are redefined with $\langle [P_\alpha (\q, \omega), P_\beta(-\q, -\omega)] \rangle$ instead of $\langle [P_\alpha (z_1', \q, \omega), P_\beta(z_2', -\q, -\omega)] \rangle$. 
If we neglect the surface charge contributions $\mathcal{C}_{bs}, \mathcal{C}_{ss}$ and impose rotational invariance in-plane, this is precisely Eq.~3 of the main text.

\subsection{Influence of Magnetic Noise}

In the main text, we quoted that the relative strength of $1/T_1$ due to magnetic noise emanating from dipoles to electrical noise is controlled by $ \mu_z^2 c_s^2/d_{\perp}^2 c^4 \sim 10^{-4} \ll 1$ for the nitrogen-vacancy center.
In this section, we derive this.

We can relate the depolarization rate due to magnetic noise $(1/T_1)_{\text{magnetic}}$ to the depolarization rate due to the electric noise $(1/T_1)_{\text{electric}}$ in the following way (we assume rotational invariance of correlations for simplicity). 
\begin{equation}
(1/T_1)_{\text{magnetic}} = \frac{(\mu_z \mu_0)^2 k_B T}{4 \pi} \int_0^\infty dq \, q \, e^{- 2 q d} \, \text{Re}[\sigma(q,\omega)] = \frac{(\mu_z \mu_0)^2 k_B T}{4 \pi} \int_0^\infty dq \, q \, e^{- 2 q d} \, \varepsilon_0 \omega \, \text{Im}[\varepsilon(\q,\omega)]
\label{eq:AppT1mag}
\end{equation}
where we have used 
\begin{equation}
    \varepsilon(\q,\omega) = 1 + \frac{i \sigma(\q,\omega)}{\varepsilon_0 \omega}
\end{equation}
Now, using the relation between the dielectric fucntion and the dynamical susceptibility $\varepsilon_0 (\varepsilon(\q,\omega) - 1) = \chi(\q,\omega)$, we note that we have $ \varepsilon_0 \text{Im}[\varepsilon(\q,\omega)] = \text{Im}[\chi(\q,\omega)]$.
Plugging the above result back into Eq.~\eqref{eq:AppT1mag}, we have
\begin{equation}
(1/T_1)_{\text{magnetic}} =  \frac{(\mu_0\mu_z)^2 k_B T \omega}{4 \pi} \int_0^\infty dq \, q \, e^{- 2 q d}  \, \text{Im}[\chi(q,\omega)] = \frac{(\mu_0 \mu_z)^2  \omega^2}{8 \pi} \left(\frac{2 k_B T}{\omega} \right) \int_0^\infty dq \, q \, e^{- 2 q d}  \, \text{Im}[\chi(q,\omega)]
\end{equation} 
Recalling that we also have
\begin{equation}
(1/T_1)_{\text{electric}} =\mu_0^2 d_\perp^2 c^4 \left( \frac{2 k_B T}{\omega} \right) \int_0^{\infty} dq \, q^3 \, e^{- 2 q d} \, \text{Im}[\chi(q,\omega)]
\end{equation}
If we consider contribution from near-resonant modes such that $\omega(q_{\rm res})$ matches the qubit splitting, then we have 
\begin{equation}
\frac{(1/T_1)_{\text{magnetic}}}{(1/T_1)_{\text{electric}}} \sim \frac{\mu_z^2 \omega^2}{d_\perp^2 c^4 q_{\rm res}^2}  = \frac{\mu_z^2 c_s^2}{d_{\perp}^2 c^4} \sim 10^{-4}
\end{equation}
where we have used $\omega = \omega(q_{\rm res}) \approx c_s q_{\rm res}$.

\section{Derivation of Spectator Mode Spectrum} \label{sec:spec}

In this section, we derive the spectrum of ``spectator'' modes that couple to both the electric field and the transverse optical phonon mode (whose dynamics are invisible to the qubit sensor), following \cite{vinogradov1992}.
As stated in the main text, these spectating modes enable probing the critical point between a paraelectric and ferroelectric at which the aforementioned transverse mode becomes soft.
The section starts by discussing the setup and approach to the derivation after which we delve into the computation of the mode spectra in both isotropic and anisotropic thin slabs.

\subsection{Setup and Main Equations}

Let us suppose that we have some dielectric material that is potentially anisotropic and that is infinite in its extent in the $xy$ plane but is finite in the $z$-direction occupying $z \in [-t, 0]$.
Above the material, it is assumed that we have the dielectric medium of the probe with dielectric constant $\varepsilon_p$.
Below the material, we will assume there is some substrate with dielectric constant $\varepsilon_s$.
Our goal will be to use Maxwell's equations to compute the spectra of modes in a material that (1) go soft at the critical point between a paraelectric and ferroelectric and (2) produce electric fields outside the material that can be sensed. 
In particular, for mode frequencies $\omega \ll qc$ (i.e. when we don't need to worry about relativistic effects), the electric field satisfies: 
\begin{equation} \label{eq-Maxwell}
    \nabla \cdot \mathbf{D} = 0 \qquad \nabla \times \mathbf{E} = 0
\end{equation}
where $\mathbf{D}$ is the electromagnetic displacement field and is related to the electric field by the dielectric tensor (or the relative permittivity tensor) by:
\begin{equation}
    D^{\alpha}_{\omega} = \varepsilon^{\alpha \beta}(\omega) E^{\beta}_{\omega}
\end{equation}
in units where the vacuum permittivity is taken to be 1. 
Assuming that dynamics of polar modes in the material take the form of a harmonic oscillator, the dielectric tensor takes the form: 
\begin{align} \label{eq-dielectric-tensor}
    \varepsilon^{\alpha \beta}(\omega(q)) &= \begin{cases} \varepsilon_s & z  > 0 \\ \left( \varepsilon_{\infty, \perp} + \frac{\Omega_{0, \perp}^2(q)}{\omega_{T, \perp}(q)^2 - \omega^2(q)} \right) \delta_{zz}^{\alpha \beta} +  \left( \varepsilon_{\infty, \parallel} + \frac{\Omega_{0, \parallel}^2}{\omega_{T, \parallel}^2(q) - \omega^2(q)} \right) \left(\delta^{\alpha \beta}_{xx} + \delta^{\alpha \beta}_{yy} \right) & z \in (0, t) \\ \varepsilon_s & z < -t
    \end{cases}
\end{align}
where $\omega(q)$ is the dispersion of a general mode in the material (which must be determined by  consistently solving Maxwell's equations), $\omega_{T, \perp/\parallel}(q)$ is the frequency of the transverse optical mode (at which, by definition, the dielectric tensor diverges), and $\Omega_{0, \perp/\parallel}$ is given by: 
\begin{equation}
    \Omega_{0, \parallel/\perp}(q) = \varepsilon_{\infty, \parallel/\perp} (\omega_{L,\parallel/\perp}^2(q) - \omega_{T, \parallel/\perp}^2(q))
\end{equation}
with $\omega_{L, \parallel/\perp}(q)$ being the dispersion of the longitudinal optical mode (at which, by definition, the dielectric tensor  vanishes).
The above choice of $\Omega_0$ makes it such that the dielectric tensor manifestly satisfies the Lydane-Sachs-Tells (LST) relation \cite{ashcroft1976solid}.
Now, let us try to determine a set of algebraic equations for the electric field given the above form of the dielectric constant.
Using translation and rotational invariance in the XY plane, we take the electric field to take the general form:
\begin{equation}
    \mathbf{E}(\mathbf{x}) = \mathbf{E}(z) e^{i q x}
\end{equation}
where we have chosen the in-plane propagation vector to be along x and positive ($q > 0$) without loss of generality due to rotational symmetry assumed in the XY plane.

Inside the probe and the substrate (i.e. for $z \notin (-t, 0)$), the dielectric constant is frequency independent and as such, Maxwell's equations [Eq.~\eqref{eq-Maxwell}] become: 
\begin{equation}
    i q  E_y = 0 \qquad \frac{d E_x}{dz} - i qE_z = 0 \qquad \frac{d E_z}{dz} + i qE_x = 0
\end{equation}
where the first two equations come from the curl equation and the last comes from divergence equation.
These equations imply that, outside of the sample and if $q = 0$,  
\begin{equation}
    E_y = 0
\end{equation}
If $q_x = 0$ (e.g., the electric field is constant in the XY plane), the $E_y$ can be anything however.
Similarly, we can solve for the other components of the electric field and we get that, again outside of the material, 
\begin{equation} \label{eq-Eoutside}
    E_x = \begin{cases} E_0 e^{- q z} & z > 0 \\ E_0' e^{q z} & z < -t \end{cases} \qquad E_z = \begin{cases} iE_0 e^{- q z} & z > 0 \\ -iE_0' e^{q z} & z < -t \end{cases}
\end{equation}
Inside the material, the equations are:
\begin{equation} \label{eq-MaxMat}
    iq_x E_y = 0 \qquad \frac{dE_x}{dz} - i q E_z = 0 \qquad \varepsilon_{\perp}(\omega)\frac{dE_z}{dz} + i q \varepsilon_{\parallel}(\omega)E_x = 0 \qquad 
\end{equation}
Once again, these equations imply that, outside of the sample and if $q = 0$, $E_y = 0$.
Finally, we remark upon the boundary conditions which are:
\begin{align} \label{eq-boundary-conditions}
    E_x(0) &= E_0 \qquad \varepsilon_\perp(\omega) E_z(0) = i\varepsilon_p E_0\\
    E_x(-t) &= E_0' \qquad \varepsilon_\perp(\omega) E_z(-t) = -i\varepsilon_s E_0'; 
\end{align}

\subsection{Spectating Modes in Isotropic Slab}

Now that we understand what the equations are and what we hope to get out of them, we can actually start solving them systematically.
Let us organize our solutions into understanding the electric field of the longitudinal modes, the transverse modes, and then finally the general Coulomb normal modes.

\subsubsection{Longitudinal Optical Modes ($\varepsilon = 0$)}

For longitudinal modes, Maxwell's equations imply that $\varepsilon = 0$.
Then from the boundary conditions of Eq.~\eqref{eq-boundary-conditions}, we have that: 
\begin{equation}
    E_0 = E_0' = 0
\end{equation}
which follows from the two equations on the right involving $\varepsilon_{\perp}(\omega)$.
Moreover, in this case, one of Maxwell's equations in Eq.~\eqref{eq-MaxMat} is satisfied trivially.
The resulting non-trivial equations are: 
\begin{equation}
    \frac{dE_x}{dz} - iq E_z = 0 \qquad E_x(0) = E_x(-t) = 0
\end{equation}
These equations are clearly underdetermined, such that $E_x(z)$ can be an arbitrary function on $[-t, 0]$ vanishing at surfaces.
As such: 
\begin{equation}
    E_x(z) = \sum_n C_n \sin\left(\frac{n \pi z}{t} \right) \qquad E_z(z) = -i \sum_n C_n \frac{\pi n}{qt} \cos\left(\frac{n \pi z}{t} \right)
\end{equation}
Interestingly, while both bulk and surface charge (determined by $4\pi \rho = \nabla \cdot \mathbf{E}$) are non-zero, the electric field vanishes identically outside of the film.

\subsubsection{Transverse Bulk Modes ($\varepsilon = \infty$)}

Since we want both the polarization and the electric field to be finite and $\mathbf{D} = \varepsilon \cdot \mathbf{E}$, for transverse modes $\mathbf{E} = 0$.
As a result, $E_0 = E_0' = 0$ from boundary conditions, implying that the electric field vanishes outside of the material.

\subsubsection{General Coulomb Normal Mode}

The most interesting case is the general coulomb normal mode where $\varepsilon \neq 0, \infty$.
The $z = 0$ boundary conditions, taken with Maxwell's equations yield:
\begin{align}
    E_x = E_0 \left( \cosh(qz) - \frac{\varepsilon_p}{\varepsilon(\omega)} \sinh(qz) \right) \qquad
    E_z = i E_0 \left(\frac{\varepsilon_p}{\varepsilon(\omega)} \cosh(qz) - \sinh(qz) \right)
\end{align}
Using the boundary conditions at $z = -t$, one gets the following condition by using $\varepsilon E_z/E_x = -i \varepsilon_s$:
\begin{equation}
    [\varepsilon^2(\omega) + \varepsilon_s \varepsilon_p] \tanh(qt) + \varepsilon(\omega) [\varepsilon_s + \varepsilon_p] = 0
\end{equation}
Solving for $\varepsilon(\omega)$ one gets:
\begin{equation}
    \varepsilon(\omega) = \frac{- (\varepsilon_s + \varepsilon_p) \pm \sqrt{(\varepsilon_s + \varepsilon_p)^2 - 4 \varepsilon_s \varepsilon_p \tanh^2(qt)}}{2 \tanh(qt)}
\end{equation}
Notice that, for the $+$ solution, $\varepsilon(\omega(q \to 0)) = 0$ because the numerator vanishes.
For the $-$ solution, $\varepsilon(\omega(q \to 0)) \to -\infty$
For our considerations, the important region is $\omega \approx \omega_T$, i.e. $|\varepsilon(\omega)| \gg 1$.
As such, we focus on the ``$-$'' solution and expand in small $q$.
This yields:
\begin{equation}
    \varepsilon(\omega) \approx - \frac{\varepsilon_s + \varepsilon_p}{qt}
\end{equation}
Using the general form of the dielectric constant Eq.~\eqref{eq-dielectric-tensor}, the dispersion becomes:
\begin{equation}
    \omega_s^2(q) = \omega_T^2(q) + qt \frac{\Omega_0^2}{\varepsilon_s + \varepsilon_p}
\label{eq:SuppXYDisp}
\end{equation}
Such a mode goes gapless at the critical point and produces an electric field outside of the material.
As such, it can be used to probe para/ferroelectric criticality.

\subsection{Spectating Modes in Anisotropic Slab}

We now turn to the anisotropic case.
Here, the equations are more interesting because in general, we cannot fine tune both $\varepsilon_{\perp} = 0$ and $\varepsilon_{\parallel} = \infty$ simultaneously.
%
%
Solving Eq.~\eqref{eq-MaxMat} with $E_x, E_z \propto e^{\lambda z}$, we get: 
\begin{equation}
    \lambda = q^2 (\varepsilon_{\parallel}/\varepsilon_{\perp})
\end{equation}
There are two cases.

\begin{description}
    \item[Case 1] (Bulk Modes) The first case is where 
    $\varepsilon_{\parallel}/\varepsilon_{\perp} < 0$.
    Here, the solutions satisfying the boundary conditions at $z = 0$ are:
    \begin{equation}
        E_x = E_0 \left( \cos\left(\alpha q z \right) - \frac{\varepsilon_p}{\alpha \varepsilon_{\perp}} \sin(\alpha q z) \right) \qquad E_z = i E_0 \left( \frac{\varepsilon_p}{\varepsilon_{\perp}} \cos(\alpha q z) - \alpha \sinh(\alpha q z) \right),
    \end{equation}
    where
    \begin{equation}
     \alpha(\omega) = \sqrt{\left|\frac{\varepsilon_\parallel(\omega)}{\varepsilon_\perp(\omega)}\right|}   
    \end{equation}
    The second boundary conditions at $z = -t$ then yield the equation: 
    \begin{equation}\label{eq-tancond}
        \tan(\alpha(\omega) qt) = \frac{\varepsilon_s + \varepsilon_p}{\alpha(\omega) \varepsilon_{\perp}(\omega) - \frac{\varepsilon_s \varepsilon_p}{\alpha(\omega) \varepsilon_{\perp}(\omega)}}
    \end{equation}
    We refer to the solutions as bulk modes because the value of the electric field is not exponentially localized to the boundaries of the slab for all mode momenta.

    \item[Case 2] (Surface-induced Modes)  $\varepsilon_{\parallel}/\varepsilon_{\perp} > 0$: In this case, solutions satisfying Maxwell's equations and the $z = 0$ boundary conditions are:
    \begin{equation}
        E_x = E_0 \left( \cosh(\alpha qz) - \frac{\varepsilon_p}{\alpha \varepsilon_{\perp}} \sinh\left(\alpha qz \right) \right) \qquad E_z = i E_0 \left( \frac{\varepsilon_p}{\varepsilon_z} \cosh\left(\alpha qz \right) - \alpha \sinh\left( \alpha q z \right) \right) 
    \end{equation}
    The second boundary condition then yields the equation:
    \begin{equation}
        \tanh\left(\alpha(\omega) qt\right) = - \frac{\varepsilon_s + \varepsilon_p}{\alpha(\omega) \varepsilon_{\perp}(\omega) + \frac{\varepsilon_s \varepsilon_p}{\alpha(\omega) \varepsilon_{\perp}(\omega)}}
    \end{equation}
    For $\alpha q t\gg1$, the electric field is localized to the boundaries of the slab at finite momenta.

    These modes are always gapped in the anisotropic case.
    To see why, note that for $\omega \to \text{min}(\omega_{T, \perp}, \omega_{T, \parallel})$, one gets that $\alpha |\varepsilon_{\perp}| \sim \pm \sqrt{\varepsilon_{\perp} \varepsilon_{\parallel}} \to \infty$.
    In this limit, a real and positive solution of $\alpha$ exists only if $\varepsilon_{\perp} < 0$ (recall that $q > 0$ and so $\tanh(\cdot) > 0$ in this case).
    This means that $\varepsilon_{\parallel} < 0$.
    This can only occur at frequencies $\omega > \text{max} \left(\omega_{T, \perp}, \omega_{T, \parallel}\right)$.
    Therefore, in the anisotropic case, the surface modes do not go soft at the critical point.
\end{description}

\subsubsection{Easy Plane (XY) Anisotropy ($\omega_{T, \parallel} < \omega_{T, \perp}$)}

Once again, we are interested in modes with $\omega \approx \omega_{T, \parallel}$.
As discusssed above, in the vicinity of this frequency, there are no surface modes but bulk modes could occur for $\omega > \omega_{T, \parallel}$.
In the regime described, 
\begin{equation}
    \tan(\alpha(\omega)q d) \approx \frac{\varepsilon_s + \varepsilon_p}{\alpha(\omega) \varepsilon_{\perp}(\omega_{T, \parallel})}
\end{equation}
One can re-write this as: 
\begin{equation}
    y\tan(y) \approx \frac{\varepsilon_s + \varepsilon_p}{\varepsilon_{\perp}(\omega_{T, \parallel})} qt 
    \label{sup:eq:xyfreq}
\end{equation}
where $y \equiv \alpha(\omega) qd$. Let us focus on the long-wavelength modes  $q t \ll 1$, so the r.h.s. of \eqref{sup:eq:xyfreq} is small. For $y\ll1$ there exists a solution:
\begin{equation}
    \alpha = \sqrt{\frac{\varepsilon_s + \varepsilon_p}{qt \varepsilon_{\perp}(\omega_{T, \perp})}} \implies \varepsilon_{\parallel}(\omega) = - \frac{\varepsilon_s + \varepsilon_p}{qt}.
\end{equation}
Other solutions are found for $y \approx \pi n$
 with $\alpha \approx \frac{\pi n}{qd}$ with $n \geq 1$.
From these solutions, the dispersions can be found:
\begin{equation}
    \omega_0^2(q) = \omega_{T, \parallel}^2 + \frac{\Omega_{0, \parallel}^2 qt}{\varepsilon_s + \varepsilon_p} \qquad \omega_n^2(q) = \omega_{T, \parallel}^2 + \frac{\Omega_{0, \parallel}^2 (qt)^2}{\varepsilon_{\perp} \pi^2 n^2} \qquad n \geq 1
\end{equation}
Note that in dropping the $q$ dependence of $\omega^2_{T, \parallel}$ and $\Omega_{0, \parallel}^2$, we are neglecting the dispersion of the transverse and longitudinal phonon.
Qualitatively, these will increase the above mode frequencies by $\sim (c_s q_z)^2 \sim (c_s \pi n/d)^2$ and additionally by $c_s^2 q^2$ due to in-plane dispersion. For a more extended discussion, see discussion on the contribution of XY bulk modes to $1/T_1$ below.
%

We conclude by discussing how these results extend to truly 2D materials.
For a monolayer, $t \to 0$, implying that only $\omega_0^2$ is a legitamate mode.
Note that, as we go towards $t \to 0$, this mode has the exact dispersion of the longitudinal optical phonon mode in 2D \cite{sohier2017}.

\subsubsection{Uniaxial (Ising) $ZZ$ Anisotropy ($\omega_{T, \perp} < \omega_{T, \parallel}$)}

In analogy to XY case, we consider bulk modes close to $\omega_{T,\perp}$.
In this case $\alpha \varepsilon_\perp\gg1$ (but $\alpha\to0$) and equations for the frequency are given by:
\begin{equation}
       y \tan y \approx \frac{\varepsilon_s+\varepsilon_p}{\varepsilon_\perp(\omega)} qt,
\end{equation}
We solve this equation in the same way as we did for the XY case in the previous section.
However, in this case $\varepsilon_\parallel\approx \varepsilon_\parallel (\omega_{T,\perp})$, while $\varepsilon_\perp\to -\infty$. Note that as $\varepsilon_\perp<0$, there is no solution for $y\to0$. The frequencies are given by:
\begin{equation}
    \begin{gathered}
        \omega_n^2(q) = \omega^2_{T,\perp}+ \frac{\Omega_{0,\perp}^2 \pi^2 n^2}{\varepsilon_\parallel (qt)^2},\; n\geq 1, qt\gg1.
    \end{gathered}
    \label{eq:Isingfreq}
\end{equation}
In this case, soft modes are at larger $q$ since they are effectively more ```transverse''.
As for the phonon mode dispersion, it can be neglected along $z$ for $\Omega_0^2\gg\varepsilon_\parallel (c_s q)^2$. In-plane dispersion is easily taken into account in this case due to translation invariance, resulting in:
\begin{equation}
    \begin{gathered}
        \omega_n^2(q) = \omega^2_{T,\perp}+ \frac{\Omega_{0,\perp}^2 \pi^2 n^2}{\varepsilon_\parallel (qt)^2} +c_s^2q^2.
    \end{gathered}
\end{equation}
As before, let us discuss how to extend these results to truly 2D materials.
The limit $t\to 0$ is now in contradiction to the assumption $qt\gg1$ taken above.
As such, we rewrite equation \eqref{eq-tancond} for $qt\ll1$:
\begin{equation}
\alpha^2 \approx \frac{\varepsilon_\parallel(\varepsilon_s+\varepsilon_p)}{\varepsilon_s\varepsilon_p qt},
\end{equation}
This implies that:
\begin{equation}
    \omega^2(q) = \omega_L^2 \left(1-\frac{\varepsilon_s\varepsilon_p qt}{\varepsilon_{\infty,\perp}(\varepsilon_s+\varepsilon_p)}\right)+c_s^2q^2,
\end{equation}
where $\omega_L \equiv \frac{\Omega_{0,\perp}}{\sqrt{\varepsilon_{\infty,\perp}}}$.
The transition occurs when the minimum of this expression reaches zero, therefore there will be in general a finite $q$ state formed.

\section{Derivation of Relaxation Rate From Spectator Modes} \label{sec:relaxspec}

Recall that in the main text, we quoted formulas for the relaxation rate of the spectator modes as a function of the mode frequency and distance to the sample.
Here, we provide a derivation of these formulas.

\subsection{Electric Field Fluctuations from Spectating Modes}

Let us recall that our formula for $1/T_1$ is given by: 
\begin{equation}
    \frac{1}{T_1} = \frac{d_{\perp}^2}{2} \coth\left( \frac{\beta \omega}{2} \right) \int \langle [E_-(t), E_+(0)] \rangle e^{i \omega t} = d_{\perp}^2 \coth\left(\frac{\beta \omega}{2}\right) \text{Im} \left[ C^R_{E_- E_+} (\omega)\right]
\end{equation}
where $C^R_{E_+ E_-} (\omega)$ is the retarded correlation functions and $\omega$ is the frequency of the qubit probe.
These electric field correlations arise from the polarization correlations in the materials.
In this section, we derive the electric field fluctuations arising from the polarization correlations in the material.
To derive the polarization correlations, let us recall that the real-size polarization operators can be expressed in terms of a normal model expansion:
\begin{equation}
\widehat{\mathbf{P}}^{\dagger}(\mathbf{r}) = \sum_{n} \mathbf{P}_n(\mathbf{r})\, \widehat{\phii}_n^{\dagger}
\end{equation}
where $\widehat{\phii}_n^{\dagger}$ are bosonic operators that satisfy the canonical commutation relations and $\mathbf{P}_n(\mathbf{r})$ is an orthogonal set of mode functions.
To properly compute the electric field fluctuations from the polarization fluctuations, we need to properly fix the normalization of $\mathbf{P}_n(\mathbf{r})$.
To do so, we consider the toy problem of non-interacting, isotropic, fully bulk problem.
In this case, the normalization of these modes can be determined by recalling that the polarization arises from phonons in the material.
From linear response theory, the correlation function of the polarization is then related to the dielectric constant.
As such:
\begin{equation}   \frac{\varepsilon(\mathbf{q}) - \varepsilon_{\infty}}{4\pi} =  \int dt\, \langle [\widehat{P}_{i, \mathbf{q}}(t), \widehat{P}_{i, -\mathbf{q}}(t) ] \rangle e^{i \omega t}   = \frac{\Omega_0^2}{4 \pi ( \omega_T^2 - \omega^2)} 
\end{equation}
where we have neglected Coulomb interaction effects.
To ensure that the above is met, we normalize:
\begin{equation}
    \int d^3 \mathbf{r}\,  P_{n}(\mathbf{r}) P_{i, m}(\mathbf{r}) = \Omega_0 \delta_{nm}
\end{equation}
which is consistent with the units of $\mathbf{P} = [\text{Energy}]/[\text{Volume}]^{1/2}$.

Normalization in hand, we can now use the mode expressions we derived for the various scenarios considered to compute the polarization correlators.
Namely, the above expressions for the mode frequencies enabled determining $E(\mathbf{r})$ with a single normalization factor $E_0$.
We use the fact that inside the material, the polarization is proportional to the electric field with $P_{\perp, \parallel} = \frac{\varepsilon_{\perp, \parallel} - 1}{4 \pi}E$ to get $E_0$.
A list $E_0(\mathbf{q})$ is reported for the different classes of modes below:
\begin{equation}
    \begin{cases}
        E_0(q) = \frac{4\pi q t}{\varepsilon_s+\varepsilon_p} \sqrt{\frac{\Omega_0}{V}}& \text{Isotropic surface modes},
        \\
       E_0^{(0)}(q) = \frac{4\pi q t}{\varepsilon_s+\varepsilon_p} \sqrt{\frac{\Omega_0}{V}};
       \;
              E_0^{(n)}(q) = \frac{4 (q t)^2}{\varepsilon_\perp \pi n^2} \sqrt{\frac{2\Omega_0}{V}};
       & \text{XY anisotropy},
        \\
     E_0^{(n)}(q) = \frac{4\pi^2 n}{\varepsilon_\parallel q t} \sqrt{\frac{2\Omega_0}{V}};  & \text{Ising 3D},
             \\
     E_0^{(n)}(q) = \frac{4\pi \varepsilon_s q t}{\varepsilon_s+\varepsilon_p} \sqrt{\frac{\Omega_0}{V}};  & \text{Ising 2D},
    \end{cases}
    \label{eq:e0}
\end{equation}
where factors of $2$ under the square root for XY and Ising cases are due to $\cos(\pi n z/t)$ [XY] or $\sin(\pi n z/t)$ [Ising 3D] profile of the field. For Ising 2D case, the field is simply constant along $z$ axis.

\subsubsection{Special Case: $1/T_1$ for the NV center}

With these expressions, we are now prepared to evaluate $1/T_1$ in terms of $E_0(\mathbf{q})$.
We will do so for the special case of the NV center to make numerical estimates shortly.
Let $\mathbf{E} = \sum_n \mathbf{E}_n \widehat{\phii}_n$.
Then:
\begin{align}
    C^R_{E_- E_+}(\omega) &= \sum_{\mathbf{q}, n, m} E_-(\mathbf{q}, n) E_+(\mathbf{q}, m) C^R_{\widehat{\phii}_n(\mathbf{q}) \widehat{\phii}_m(\mathbf{q})}(\omega) \\
    &= \sum_{\mathbf{q}, n} \left( \frac{}{}|E_{X}(\mathbf{q}, n)|^2 +  |E_{Y}(\mathbf{q}, n)|^2  \right) C^R_{\widehat{\phii}_n(\mathbf{q}) \widehat{\phii}_n(\mathbf{q})}(\omega)
\end{align}
where $X$ and $Y$ are the axes of the NV.
Note that the field outside the film has the same amplitude along $x$ and $z$, see Eq. \eqref{eq-Eoutside}. In the calculations above, $x$ axis is along ${\bf q}$ whereas in the expression, the axes are fixed. Therefore, averaging over the direction of ${\bf q}$ in the film plane will reduce the overall contribution of the in-plane component of electric field fluctuations. On the other hand, the $z$ component does not depend on ${\bf q}$. 

Thus, the result depends on the orientation of the NV center's $\tilde{z}$ axis with respect to the $z$ axis of the film described by polar angle $\theta$. We can average over the in-plane azimuthal angle of ${\bf q}$ then to get:
\begin{equation}
\begin{gathered}
        C^R_{E_- E_+}(\omega) = \left(1/2+1/2\cos^2\theta+\sin^2\theta\right)\sum_{{\bf q},n} |E_0({\bf q},n)|^2 e^{-2qd}
    C^R_{\widehat{\phii}_n(\mathbf{q}) \widehat{\phii}_n(\mathbf{q})}(\omega),
\end{gathered}
\label{eq:angav}
\end{equation}
where $d$ is the NV center height above the sample and
\begin{equation}
 \chi^R_{\hat{P}_{{\bf q},n},\hat{P}_{{\bf q},n}}
  = 
  \frac{1}{4\pi}
  \frac{\Omega_0}{\omega^2({\bf q},n)-(\omega+i\delta)^2}.
\end{equation}
The largest prefactor in \eqref{eq:angav} is obtained when the NV axis is in the plane (such that one of the perpendicular directions is along $z$ axis).
For positive frequencies, we can evaluate the imaginary part to get:
\begin{equation}
      \frac{1}{T_1} 
    =
     \frac{3d_\perp^2}{16}
     \coth\left( \frac{\beta \omega}{2} \right)
\sum_{n} \int \frac{d^2 q}{(2\pi)^2} |E_0({\bf q},n)|^2e^{-2qd}
   \frac{\Omega_0 \delta(\omega-\omega({\bf q},n))}{\omega({\bf q},n)}.
   \label{eq:result}
\end{equation}

We now apply \eqref{eq:result} to the cases considered above.

\subsection{XY and isotropic case}
\subsubsection{Surface-induced spectating modes}
We consider the system at the critical point, such that $\omega(q)=\sqrt{qt \frac{\Omega_0^2}{\varepsilon_s+\varepsilon_p}}$ (assuming $qd\ll1;\; qt\ll1$):

\begin{equation}
\begin{gathered}
\frac{1}{T_1} 
    =
\frac{3d_\perp^2}{16}
     \coth\left( \frac{\beta \omega}{2} \right)
\int \frac{q dq }{2\pi} \left(\frac{4\pi q t}{\varepsilon_s+\varepsilon_p}\right)^2 \frac{\Omega_0^2}{t}\frac{\delta(\omega(q)-\omega)}{\omega(q)} = 
\\
= \frac{3\pi d_\perp^2 (\varepsilon_s+\varepsilon_p)^2}{t^3} \coth(\omega/2T) \frac{\omega^6}{\Omega_0^6}.
\end{gathered}
   \label{eq:isotrt1}
\end{equation}

To get an estimate for $1/T_1$ for NV center, we use $d_{\perp} = 17\ \mathrm{ Hz}\cdot\mathrm{cm/V}$ \cite{VanOort1990}. One can compare this value with Bohr magneton in Gaussian units $\mu_B = \frac{e \hbar}{2 m_e c} = 4.7\cdot 10^3\mathrm{ Hz}\cdot\mathrm{cm/V}$, such that $d_{\perp} =3.6 \cdot 10^{-3} \mu_B$. Introducing this value and using atomic units, one gets
\begin{equation}
   \frac{1}{T_1} [MHz] =  10.7 \cdot(\varepsilon_s+\varepsilon_p)^2 (a_B/d)^3 \coth(\omega/2T) \frac{\omega^6}{\Omega_0^6},
\end{equation}
where we used $\mu_B^2/a_B^3 = \frac{E_H^2}{4 m_e c^2} \approx 0.087$ THz, $E_H$ being the Hartree energy.
With $\Omega_0$ value for STO around $200 meV\approx 50$ THz, minimal thickness around $t=1$ nm and frequencies $\omega\ll T$ we get
\begin{equation}
       \frac{1}{T_1}[ Hz]  \approx 2 \cdot 10^{-7} (\varepsilon_s+\varepsilon_p)^2 T \omega^5[THz].
\end{equation}
The only way to get appreciable result is to have a substrate with large dielectric constant is to employ a material with large dielectric constant as a substrate. For example, using SrTiO$_3$ as a substrate ($\varepsilon\sim 10^4$) will allow to get $1/T_1\sim Hz$ for frequencies of order $1$  THz.






\subsubsection{Bulk mode contribution}

$\bullet$ XY anisotropy
\\ \indent
In this case, there is a contribution of a surface-induced mode, identical to Eq. \eqref{eq:isotrt1}, and, in addition to that, a contribution to electric field from the bulk transverse-like modes. Physically, these are modes polarized predominantly along $x$ with momentum predominantly along $z$ ($q_z\gg q$). In the bulk limit, there is one transverse and one longitudinal mode with such momentum. The breaking of the rotational symmetry by the film geometry mixes them. However, in the isotropic case, the diverging dielectric constant in all directions still screens the resulting electric field. The finite value of dielectric constant along $z$ in the anisotropic case, implies that polarization oscillations are accompanied by a finite electric field: $E_z\sim 4 \pi P/(\varepsilon_\perp-1)$.
\\ \indent
Here an important role may be played by the phonon dispersion. Including dispersion effects in the in-plane direction in the current formalism is straightforward: the dispersion of the n-th mode is simply given by $\tilde{c}_s(n) q$, where
\begin{equation}
	\tilde{c}_s(n)  = \sqrt{c_s^2+ \frac{\Omega_{0,\parallel}^2}{\varepsilon_\perp (\pi n /t)^2}}
\end{equation}
\indent
It is not so for the $z$ direction. Therefore, we estimate when the effects of $z$-axis dispersion become important and analyze the limiting cases. For the bulk modes $\alpha\approx \frac{\pi n}{q t}$ and the polarization profile from above follows to simply consist of harmonics  $\vec{P} \propto \cos (\pi n z/t), \sin(\pi n z/t)$. Assuming that $t\gg a_0$, we can use harmonic approximation for the dispersion $E_{disp,\perp} = \int_{-t}^0 c_{s,\perp}^2 (\partial_z \vec{P})^2 dz = c_{s,\perp}^2\vec{P}\partial_z \vec{P} |_{-t}^0  - \int_{-t}^0 c_{s,\perp}^2 \vec{P} \partial_{zz} \vec{P}   dz$, where we used integration by parts/Stokes' theorem. The first term depends on the boundary conditions for polarization and is not universal, but the second one simply returns $(c_{s,\perp}\pi n/t)^2$, as $\cos (\pi n z/t), \sin(\pi n z/d)$ are eigenfunctions of $ \partial_{zz}$. For large $n$, the first term is expected to grow not faster than $n$ and thus can be neglected. Furthermore, for $\varepsilon_\parallel \to \infty$ the edge term is suppressed as $1/\sqrt{\varepsilon_\parallel}$ and can be thus neglected for all $n$. 
\\ \indent
Thus, for $n\gg1$ we can approximate the full dispersion by:
\begin{equation}
\omega(q,n)_{n\gg1} =  \tilde{c}_s^2 q^2 + (c_{s,\perp} \pi n/t)^2.
\end{equation}
For fixed $\omega$ this implies that mode contributing to $1/T_1$ will have $q = q_n = \sqrt{(\omega^2-(c_{s,\perp} \pi n / t)^2)}/\tilde{c}_s< \omega_{c_s}$ which implies a cutoff $n\lesssim n_\perp^{\max}$, where
\begin{equation}
	n_\perp^{\max} = \frac{\omega}{\pi c_{s,\perp}/t}.
 \label{eq:nperp}
\end{equation}
\\ \indent
On the other hand, if we neglect the $z$ axis dispersion, the sum over $n$ is cut off by the exponential factor $e^{-q d}$. Since $\tilde{c}_s(n)$ decreases with $n$, $q_n$ grows with $n$. This implies that there is a characteristic value $n_\parallel^{max}: q_{n_\parallel^{max}} = 1/d$, such that the contributions of modes with $n>n_\parallel^{max}$ to $1/T_1$ are suppressed:
\begin{equation}
	n_\parallel^{max} = \frac{t}{\pi d} \frac{\Omega_{0,\parallel}/\sqrt{\varepsilon_\perp}}{\sqrt{\omega^2-(c_s/d)^2}}
 \label{eq:npar}
\end{equation}
\\\indent
We will consider now two limits, depending on the relation between $n_\perp^{max}$ and $ n_\parallel^{max}$.
\\
$\bullet$ For $n_\perp^{max}\ll n_\parallel^{max}$, $q_n\ll 1/d$ for relevant values of $n$ and thus the exponential in the sum can be neglected.  Comparing Eq. \eqref{eq:nperp} and Eq. \eqref{eq:npar} one finds that $n_\perp^{max}\ll n_\parallel^{max}$ corresponds to either $\omega\ll c_s/d$ or $c_s/d \ll \omega \ll \sqrt{(\Omega_{0,\parallel}/\sqrt{\varepsilon_\perp})(c_{s,\perp}/d)}$. The expression for $1/T_1$ is:
\begin{equation}
\begin{gathered}
\frac{1}{T_1} 
    \approx
    \frac{3d_\perp^2}{16}
     \coth\left( \frac{\beta \omega}{2} \right)
     \sum_{n=1}^{\omega/(\pi c_{s,\perp}/ t)}
\int \frac{q dq }{2\pi} \left(\frac{4 \pi \omega^2(q,n)}{\Omega_{0,\parallel}^2}\right)^2 \frac{2\Omega_{0,\parallel}^2}{t}\frac{\delta(\omega(q,n)-\omega)}{\omega(q)}
=
\\
=
\frac{3 \pi d_\perp^2 \omega^4}{t \Omega_{0,\parallel}^2}
     \coth\left( \frac{\beta \omega}{2} \right)
     \sum_{n=1}^{\omega/(\pi c_{s,\perp}/ t)}
\frac{1}{\tilde{c}_s^2(n)}
\approx \left/
\omega\ll 
\frac{\Omega_{0,\parallel} (c_{s,\perp}/c_s)}{\sqrt{\varepsilon_\perp}}\right/\approx
\\
\approx 
d_\perp^2 \varepsilon_\perp 
\left(\frac{\omega}{\Omega_{0,\parallel}}\right)^4
\left(\frac{\omega}{c_{s,\perp}}\right)^3
     \coth\left( \frac{\beta \omega}{2} \right).
\end{gathered}
   \label{eq:xycase2}
\end{equation}
\\
$\bullet$  $n_\perp^{max}\gg n_\parallel^{max}$, which requires $c_s/d , \sqrt{(\Omega_{0,\parallel}/\sqrt{\varepsilon_\perp})(c_{s,\perp}/d)} \ll \omega$. This limit can be realized for weak anisotropy and NV center sufficiently far from the surface. For example, for a micron-thick STO ($c_s=6.6\cdot 10^3$ m/s, so $\omega\gg 6.6. THz/ h[nm]$) frequencies above GHz should be ok.
In the case $\omega \gg \frac{c_s}{d}, \sqrt{\frac{c_s}{d}\frac{\Omega_0}{\varepsilon_\perp}}$ one can neglect $c_s$ altogether. Then, using expression \eqref{eq:result} we can get the following expression for the contribution to the relaxation rate, ignoring the phonon dispersion $c_s q$:
\begin{equation}
\begin{gathered}
\frac{1}{T_1} 
    =\frac{3d_\perp^2}{16}
     \coth\left( \frac{\beta \omega}{2} \right)
     \sum_n
\int \frac{q dq }{2\pi} \left(\frac{4 (q t)^2}{\varepsilon_\perp \pi n^2}\right)^2 \frac{2\Omega_0^2}{t}\frac{\delta(\omega(q)-\omega)}{\omega(q)} e^{-2 q d}
\\
=
\frac{3d_\perp^2 \pi}{t^3}
\coth\left( \frac{\beta \omega}{2} \right)
\left(\frac{\omega}{\Omega_0}\right)^4
\sum_n \varepsilon_\perp \pi^2 n^2 \exp\left(-\sqrt{\varepsilon_\perp}\frac{2 d}{t} \frac{\omega}{\Omega_0} n\right)=
\\
=
\frac{3d_\perp^2 \pi^3 \varepsilon_\perp}{t^3}
\coth\left( \frac{\beta \omega}{2} \right)
\left(\frac{\omega}{\Omega_0}\right)^4
\frac{e^a(1+e^a)}{(e^a-1)^3},
\end{gathered}
   \label{eq:xycase1}
\end{equation}
where $a=\sqrt{\varepsilon_\perp}\frac{2 d}{t} \frac{\omega}{\Omega_0}$. For large $a$ it decays exponentially, however, given that $\omega\ll \Omega_0$ the opposite limit $a\ll1$ is more likely. For that case:
\begin{equation}
\begin{gathered}
\frac{1}{T_1} 
\approx
     \frac{3d_\perp^2 \pi^3 }{4 \sqrt{\varepsilon_\perp}d^3}
     \coth\left( \frac{\beta \omega}{2} \right)
     \frac{\omega}{\Omega_0}
          \ll
    \frac{3d_\perp^2 \pi^3 \varepsilon_\perp }{4 t^3}
     \coth\left( \frac{\beta \omega}{2} \right)
     \left(\frac{\omega}{\Omega_0}\right)^4
     \left(\frac{\omega}{c_{s,\perp}/t}\right)^3
\end{gathered}
   \label{eq:xycase1limit}
\end{equation}
\\
We can now compare the contribution of bulk modes and the surface-induced mode. From the above it follows that the case \eqref{eq:xycase1} results in a larger result, so we focus on comparing it with \eqref{eq:isotrt1}. We have:
\begin{equation}
    \left(\frac{1}{T_1}\right)_{bulk} \sim \left(\frac{1}{T_1}\right)_{2D\;mode} \frac{\varepsilon_\perp}{(\varepsilon_s+\varepsilon_p)^2}     \frac{\omega \Omega_0^2}{(c_{s,\perp}/t)^3}.
\end{equation}
Typical $c_s$ value is $10^3m/s$, so $c_s/t\sim 1 THz/t[nm]$, making the latter factor in the expression to be around 10 for $\omega\sim 100$ GHz and thickness $t=10$ nm. Note that $\left(\frac{1}{T_1}\right)_{bulk}$ is thickness-independent, so for fair comparison with the 2D contribution we take a minimal thickness.
If a regular substrate is used, $(\varepsilon_s+\varepsilon_p)\sim1$, the bulk contribution is larger. However, with an STO substrate, $\frac{\varepsilon_\perp}{(\varepsilon_s+\varepsilon_p)^2}$ can be of the order $10^{-4}$ or smaller, making the 2D mode contribution dominant. Thus, the most favorable case for the observation of soft fluctations of a bulk ferroelectric would be still to use the 2D mode on a high$-\varepsilon$ substrate.

$\bullet$ Ising anisotropy

Let us first analyze the collective mode dispersion, Eq. \eqref{eq:Isingfreq}. It has a minimum at a finite $q=q_0$ and can be written in its vicinity as:
\begin{equation}
    \omega^2({\bf q}) = \omega_0^2 + c'^2 (q-q_0)^2
\end{equation}
where 
\begin{equation}
\begin{gathered}
\omega_0^2 = 
    \begin{cases}
      \omega_T^2+2 \frac{\Omega_{0,\perp}}{\sqrt{\varepsilon_\parallel}} \frac{\pi n c_s}{t}&  3D,
        \\
     \omega_L^2 - \left(\frac{\varepsilon_s\varepsilon_p}{\varepsilon_\infty(\varepsilon_s+\varepsilon_p)}\right)^2
     \frac{\omega_L^4}{4 (c_s/t)^2} & 2D
    \end{cases}
\\
c' = 
    \begin{cases}
      2c_s&  3D,
        \\
    c_s & 2D
    \end{cases}
\\
q_0 = 
    \begin{cases}
     \frac{1}{c_s}\sqrt{\frac{\Omega_{0,\perp}}{\sqrt{\varepsilon_\parallel}}
     \frac{\pi c_s n}{t}
     }&  3D,
        \\
    \frac{\omega_L^2 t \varepsilon_s\varepsilon_p}{2c_s^2(\varepsilon_s+\varepsilon_p)} & 2D
    \end{cases}
\end{gathered}
\end{equation}
\\
At criticality, $\omega_0\to 0$. Moreover, assuming $2 \frac{\Omega_{0,\perp}}{\sqrt{\varepsilon_\parallel}} \frac{\pi c_s}{t}\gg\omega^2$ we can neglect the $n>1$ modes for the 3D case. Combining \eqref{eq:result} and \eqref{eq:e0}, we get (note that delta function gives two contributions at $q=q_0\pm\omega/c'$):
\begin{equation}
    \begin{gathered}
           \left(\frac{1}{T_1} \right)_{Ising}
    =
     \frac{3d_\perp^2}{16}
     \coth\left( \frac{\beta \omega}{2} \right)
\int \frac{d^2 q}{(2\pi)^2} |E_0({\bf q},n)|^2e^{-2qd}
   \frac{\Omega_0 \delta(\omega-\omega({\bf q},n))}{\omega({\bf q},n)}
   \approx
   \\
   \approx /\omega\ll c' q_0/\approx
\frac{3d_\perp^2}{16}
\coth\left( \frac{\beta \omega}{2} \right)
|E_0(q_0,n)|^2e^{-2q_0d}
\frac{2 q_0\Omega_0}{2\pi c' \omega}
=
\\
=
    \begin{cases}
      \frac{3 d_\perp^2 \pi^3 \Omega_{0,\perp}^2}{\varepsilon_\parallel^2 t^3 c_sq_0 \omega}
    \coth\left( \frac{\beta \omega}{2} \right)  e^{-2q_0d}
      &  3D,
        \\
      \frac{3 d_\perp^2 \pi \Omega_{0,\perp}^2 \varepsilon_s^2 q_0^3}{(\varepsilon_s+\varepsilon_p)^2 (c_s/t) \omega}
   \coth\left( \frac{\beta \omega}{2} \right) e^{-2q_0d}& 2D
    \end{cases},
    \end{gathered}
\end{equation}
where where $\omega_L \equiv \frac{\Omega_{0,\perp}}{\sqrt{\varepsilon_{\infty,\perp}}}$.
Both results are dramatically larger then the one for the isotropic or XY case. The only issue is the exponential factor, suppressing stray field for large $q_0$ (such as the case for antiferroelectrics). The other thing to keep track of is that the above expression was derived for $q_0t\gg1$ for 3D case and $q_0t\ll 1$ for the 2D case. As a function of thickness, the result is $\propto 1/t^3$ in the former case and $\propto t$ in the latter one. This suggests that there is an optimal regime $t q_0 \sim 1$, where the signal is maximized.

We can also estimate $1/T_1$ for 2D case using $t\Omega_0^2 = 10^{-2}- 10^{-4} $ angstrom eV$^2$, \cite{sohier2017} taking $c_s\sim 10^4$ m/s (an overestimate likely) to get $1/T_1[\mathrm{Hz}] \sim 10^7\cdot (q_0 a_B)^3 \coth\left( \frac{\beta \omega}{2} \right) e^{-2q_0d}$. Taking $T\sim 50 $ K (1 THz) and $\omega\sim1$ GHz and $q_0^{-1}\sim 100$ nm (and $q_0d\ll1$) we get $1/T_1 \sim 10^2- 10^4$ Hz.

\section{Derivation of $1/T_1$ Scaling across Para-to-Ferroelectric Phase Transitions}\label{sec:phasetransition}

In this section, we provide additional details regarding how qubit sensors can shed light on paraelectric-to-ferroelectric phase transitions.
We first consider a thermal (classical) phase transition where we detail the derivation of the scaling theory of the main text.
We conclude by doing the same for the quantum case.

\subsection{Derivation of $1/T_1$ near thermal ferroelectric transitions: Single mode approximation and scaling theory}

We discuss the behavior of $1/T_1$ near a thermal critical point for a clean ferroelectric material.
Since we want to go beyond the single-mode approximation for dynamical polarization correlations in the sample of interest, we use the formulation for $1/T_1$ derived using Maxwell's equations for propagating the stray electric fields.
\beq
\frac{1}{T_1} = \frac{d_\perp^2 \mu_0^2}{2} \coth\left( \frac{\beta \omega_q}{2} \right) \int \frac{d^2q}{(2 \pi)^2} F(d,q) C(\omega_q, q)
\eeq
where $F(d,q) = \sum_{j=0}^{N_\ell-1} q^2 e^{- 2 q (d + j w)}/16$ is a filter function for a $N_\ell$-layered system with an inter-layer distance of $w$, and $C(\omega_q, q)$ corresponds to the imaginary part of longitudinal polarization correlations that contribute to electrical noise.

In the single-mode approximation, the imaginary part of the polarization correlations is a Dirac-delta function at the mode frequency $\omega = \omega(q)$, i.e.,
\beq
C(\omega,q) \sim \text{Im}\left[ \frac{1}{(\omega + i 0^+)^2 - \omega^2(q)} \right] = \frac{1}{2 \omega(q)} \left[ \delta(\omega - \omega(q)) - \delta(\omega + \omega(q)) \right]
\eeq
If we further consider the physically relevant scenario where $\omega(q) = \omega$ has a unique solution as a function of $q = |\q|$, we can write
\beq
\delta(\omega - \omega(q)) = \frac{\delta(q - q_0)}{|\frac{d\omega}{dq}|_{q = q_0}} = \frac{\delta(q - q_0)}{|v_g(q_0)|}
\eeq
where $v_g(q_0) = \partial_q \omega(q)|_{q = q_0}$ is the group velocity of the excitation at momentum $q_0$.
Therefore, we have (assuming the 2d limit $w N \ll d$) 
\beq
\frac{1}{T_1} \propto \frac{d_\perp^2 \mu_0^2}{2} \coth\left( \frac{\beta \omega_q}{2} \right) \frac{N_\ell}{32 \pi} \int_0^\infty dq \, q^3 e^{- 2 q d}  \frac{\delta(q - q_0)}{2 \omega(q)|\frac{d\omega}{dq}|_{q = q_0}} = \frac{d_\perp^2 \mu_0^2 N_\ell}{128 \pi} \coth\left( \frac{\beta \omega_q}{2} \right) \frac{q_0^3 e^{- 2 q_0 d}}{\omega(q_0) |v_g(q_0)|}
\eeq
In this limit, therefore, the physics is entirely determined by the dispersion of the mode. 
Generally, for the polarization mode of interest for the easy plane ferroelectric, we derived the following dispersion for the generalized Coulomb nomral mode in the previous section (see Eq.~\eqref{eq:SuppXYDisp})
\beq
\omega^2(\q) = \omega_{T}^2(\q) + a |\q|  = \omega_{0,T}^2 + c_s^2 \q^2 + a |\q|, \text{ where } a = \frac{\varepsilon_\infty \omega_{L}^2 t}{\varepsilon_s + \varepsilon_p}
\eeq
where $\omega_{T}(\q)$ is the transverse mode frequency, and the non-analytic term proportional to $|\q|$ originates from long-range Coulomb interaction in two-dimensions. 
For small momenta, we have $\omega(q)^2 \approx \omega_{0,T}^2 + a |\q|$, such that $\omega(q) \partial_q \omega(q) = a$.
In this case, we get $1/T_1 \sim \coth(\beta \omega_q/2) q_0^3 e^{-2 q_0 d}$.
One may extract $q_0$ from the depolarization rate, and hence the gap of the transverse optical phonon mode $\omega_{0,T} = \sqrt{\omega_q^2 - aq_0}$. 
The critical point is indicated by the gap going to zero, which corresponds to the largest depolarization rate $1/T_1$. 

The presence of short-range interactions between critical modes can lead to dynamical polarization correlations distinct from mean-field results. 
Specifically, the pole in the correlation function $C(\omega, q)$ may be replaced by a branch cut, and the single mode approximation ceases to be valid.
Going beyond the previously discussed mean-field result, one may derive more general results the following argument.
The mean-field result for the polarization correlations is set by the inverse propagator for the relevant spectator phonon mode, i.e., 
\beq
C^{-1}(\omega,\q) \sim G^{-1}(\omega, \q) = \omega^2 - \omega(\q)^2 = \underbrace{\omega^2 - \omega_{T}^2(\q)}_{G_{T}^{-1}(\omega,\q)} - \underbrace{a |\q|}_{\Sigma_C(\omega,\q)}
\eeq
where $G^{-1}_{T}$ is the propagator for the critical transverse mode, while $\Sigma_C(\omega,\q) = a |\q|$ denotes the (instantaneous) Coulomb interaction correction owing to the fact that the longitudinal mode creates charge imbalances. We expect that this form continues to hold beyond the mean-field treatment, with the mean-field expression for $G^{-1}_{T}$ being replaced by its critical scaling form, since the single spectator mode is not expected to affect the bulk critical behavior significantly. In particular, if we break down $G^{-1}_{T} = \omega^2 - \omega_{T}^2(\q)+\Sigma_{cr}$, we ignore the effects of the spectating mode on $\Sigma_{cr}$. For a bulk system, this would be neglecting the effect of a single mode, whereas the number of modes contributing to  $\Sigma_{cr}$ is proportional to thickness and, therefore, macroscopic. In that case, $\Sigma_{cr}$ for the spectating mode arises due to its coupling to bulk modes and is this a simple self-energy insertion on par with $\Sigma_C$. The total self-energy for the spectating mode is then just $\Sigma_{cr}+\Sigma_C$.



For $G_T$ near transition we assume the scaling form:
\beq
G^{-1}_{T}(\omega, \q) \sim q^{2 - \eta} \Psi_{sc}\left( \frac{\omega}{(c_s q)^z}, \frac{\omega}{\omega_{0,T}} \right), \text{ where } \omega_{0,T} \sim |T - T_c|^{z \nu} 
\eeq
Here $\Psi_{sc}$ is a dimensionless scaling function that satisfies $\Psi_{sc}(0,0) \to 1$ (by definition of the anomalous critical exponent $\eta$).
Consequently, the imaginary part also obeys the same scaling.
Therefore, we may write the scaling form of the full correlation function as:
\beq
C(\omega, \q) &=& \text{Im} \left[ \frac{1}{q^{2 - \eta} \Psi_{sc}\left( \frac{\omega}{(c_s q)^z}, \frac{\omega}{\omega_{0,T}} \right) + \Sigma_C(\omega, \q)} \right] \nn
&=& \frac{q^{2-\eta}\text{Im}\left[\Psi_{sc}\left( \frac{\omega}{(c_s q)^z}, \frac{\omega}{\omega_{0,T}} \right) \right]}{\left(  \text{Im}\left[ q^{2-\eta}\Psi_{sc}\left( \frac{\omega}{(c_s q)^z}, \frac{\omega}{\omega_{0,T}} \right) \right] \right)^2 + \left( \text{Re}\left[ q^{2-\eta}\Psi_{sc}\left( \frac{\omega}{(c_s q)^z}, \frac{\omega}{\omega_{0,T}} \right) \right] + \Sigma_C(\omega,\q)\right)^2}
\eeq
where we have used that the Coulomb self-energy $\Sigma_C(\omega, \q) = a |\q|$ is real. 
Expressing the correlation function in this scaling form sets the stage to investigate two distinct scaling limits to extract critical exponents. 

First, let us consider the $\q \to 0$ limit, when we can re-write the scaling function as 
\beq
G^{-1}_{T}(\omega, \q \to 0) = (\omega_{0,T})^{(2-\eta)/z} \Psi_{sc}\left( \frac{\omega}{\omega_{0,T}} \right)
\eeq
As $\omega \to 0$, we expect the imaginary part of $G^{-1}_{T}(\omega,0)$, which is odd in $\omega$ due to causality, to vanish, while the real part, which is even in $\omega$, remains finite. 
This implies that 
\beq \text{Im}\left[\Psi_{sc}\left( \frac{\omega}{\omega_{0,T}} \right) \right] \xrightarrow{\omega \to 0} \frac{\omega}{\omega_{0,T}} , ~~~  \text{Re}\left[\Psi_{sc}\left( \frac{\omega}{\omega_{0,T}} \right) \right] \xrightarrow{\omega \to 0} \rm{const.}
\eeq
Therefore, in the limit where the qubit-sample distance is large compared to the correlation length so that we are sensitive to the $q \to 0$ limit of the correlation function, we may write the following scaling form for the relaxation time.
\beq
\frac{1}{T_1} \propto \coth\left( \frac{\beta \omega}{2} \right) \left( \int_0^{\infty} dq \, q^3 e^{-2qd} \right) \frac{1}{\omega_0^{(2-\eta)/z}} \left( \frac{\omega}{\omega_{0,T}} \right) \bar{\Psi}\left(\frac{\omega}{\omega_{0,T}}  \right) \xrightarrow{\omega \ll k_B T} \frac{2 T}{ \omega_{0,T}^{(z + 2-\eta)/z} d^4} \bar{\Psi}\left(\frac{\omega}{\omega_{0,T}}  \right)
\eeq
Note that this is consistent with our single mode approximation/mean-field results, which simply correspond to $z = 1$, $\eta = 0$.
Specifically, for the imaginary part, we have
\beq
\text{Im}[G_{T}(\omega, \q \to 0)] = \frac{1}{\omega_{0,T}^2} \frac{\pi}{2}\left( \frac{\omega}{ \omega_{0,T}}\right) \left[\delta\left(1 - \frac{\omega}{\omega_{0,T}}\right) + \delta\left(1 + \frac{\omega}{\omega_{0,T}}\right) \right]
\eeq
which is an odd-function in $\omega$, but with delta-function peaks which will be smoothened out beyond mean-field theory. 

We now turn to the limit where $\omega \to 0$, but $q \neq 0$. 
In this case, it is convenient to focus on the critical point $\omega_{0,T} = 0$, as in this limit the scaling function only depends on $\omega$ and $q$, and therefore we can extract useful information about scaling at this point. 
Since the imaginary part of the scaling function has be odd in frequency $\omega$, we can appeal to scaling to entirely determine its form at low-frequencies.
\beq
\text{Im}\left[\Psi_{sc} \left( \frac{\omega}{(c_s q)^z} \right) \right] = \frac{\omega}{(c_s q)^z} \tilde{\Psi}\left( \frac{\omega}{(c_s q)^z} \right), \text{ where } \tilde{\Psi}(x) \xrightarrow{x \to 0} \text{const.}
\eeq
Therefore, in the low-frequency limit, the numerator of $C(\omega,\q)$ scales as $\omega/q^z$. 
Since $\eta \ll 1$, the corrections proportional to $q^{2 - \eta}$ in the denominator will be dominated by the Coulomb term $\Sigma_C(\q,0) = a |\q|$, so the correlation function take the following form:
\beq
C(\omega,\q) = \frac{ q^{2-\eta} \left( \omega/q^z \right)}{(\Sigma_C(\q,0))^2} = \frac{\omega q^{2-\eta - z}}{a^2 q^2} \sim \omega q^{-\eta -z}
\eeq
Therefore, in this limit, we have the following scaling of the relaxation time.
\beq
\frac{1}{T_1} \propto \coth\left( \frac{\beta \omega}{2} \right) \omega \left( \int_0^{\infty} dq \, q^{3- \eta - z} e^{-2qd} \right) \xrightarrow{\omega \ll k_B T} \frac{2 T}{d^{4 - \eta - z}} 
\eeq


\subsection{Derivation of $1/T_1$ near quantum ferroelectric transitions}
Let us apply the same approximation in the quantum critical case, where the gap goes to zero as $\omega_T \sim |\lambda - \lambda_c|^{z\nu}$ as we approach the critical point $\lambda \to \lambda_c$.
Therefore, extracting the gap directly tells us about the exponents $z \nu$.
Specifically, the inverse Green's function $G_T^{-1}$ for the quantum critical mode is given by the following scaling function near the quantum critical point
\beq
G^{-1}_{T}(\omega, \q) \sim q^{2 - \eta} \Phi_{sc}\left( \frac{\omega}{(c_s q)^z}, \frac{\omega}{T}, \frac{\omega_{0,T}}{T} \right), \text{ where } \omega_{0,T} \sim |\lambda - \lambda_c|^{z \nu}
\eeq
where $\Phi_{sc}$ is a dimensionless scaling function.
In analogy with the classical/thermal phase transition, we can write the scaling form of the correlation function $C(\omega,\q)$ as follows:
\beq
C(\omega, \q) = \frac{q^{2-\eta}\text{Im}\left[\Phi_{sc}\left( \frac{\omega}{(c_s q)^z}, \frac{\omega}{T}, \frac{\omega}{\omega_{0,T}} \right) \right]}{\left(  \text{Im}\left[ q^{2-\eta}\Phi_{sc}\left( \frac{\omega}{(c_s q)^z}, \frac{\omega}{T}, \frac{\omega}{\omega_{0,T}} \right) \right] \right)^2 + \left( \text{Re}\left[ q^{2-\eta}\Phi_{sc}\left( \frac{\omega}{(c_s q)^z}, \frac{\omega}{T}, \frac{\omega}{\omega_{0,T}} \right) \right] + \Sigma_C(\omega,\q)\right)^2}
\label{eq:ScalingQ}
\eeq
Once again, we can focus on two separate regimes. 
First, we consider tuning to the critical point $\lambda = \lambda_c$, such that $\omega_{0,T} = 0$, and look at the small $\omega$ limit. 
Since $\eta \ll 1$, the denominator in Eq.~\eqref{eq:ScalingQ} is dominated by the Coulomb self-energy $\Sigma_C(\q, \omega) = a |\q|$, and can be approximated as $[\Sigma_C(\q,\omega)]]^2 = a^2 q^2$.
At the same time, the response function being odd in $\omega$ indicates that we can write 
\beq
\text{Im}\left[\Phi_{sc}\left( \frac{\omega}{(c_s q)^z}, \frac{\omega}{T},0 \right) \right] \xrightarrow{\omega \to 0} \left( \frac{\omega}{T} \right) \bar{\Phi}_{sc}\left( \frac{c_s q}{T^{1/z}} \right)
\eeq
This leads to the following scaling of $1/T_1$ as a function of temperature $T$ and qubit-sample distance $d$:
\beq
\frac{1}{T_1} \approx \coth\left( \frac{\beta \omega}{2} \right) \left( \frac{\omega}{T} \right) \int dq \, q^3 e^{-2qd} \left(\frac{q^{2-\eta}}{q^2} \right) \bar{\Phi}_{sc} 
 \left( \frac{c_s q}{T^{1/z}} \right) \xrightarrow{\omega \ll k_B T} \frac{1}{d^{4 - \eta}} \Phi(d T^{1/z})
\eeq
Therefore, by performing a scaling collapse as a function of $T$ and $d$, one can extract the scaling exponents $\eta$ and $z$.

Next, we consider tuning away from the critical point, such that we can consider the scaling function in the low-frequency and small-momentum limit. 
At higher temperatures $T \gg \omega, c_s q$, we expect that the scaling function $\Phi_{sc}$ is not divergent, and its form is fixed by scaling. 
\beq
G^{-1}_{T}(\omega, \q \to 0) \sim T^{(2-\eta)/z} \phi_{sc}^{-1}\left( \frac{\omega_{0,T}}{T}, \frac{\omega}{T} \right)
\eeq
For $1/T_1$, we need the imaginary part of the correlation function, which is always odd in $\omega$. 
Therefore, we expect that in the low-frequency limit $\omega \ll k_B T$, it takes the following form. 
\beq
C(\omega \to 0, q \to 0) \sim T^{(-2+\eta)/z} \left( \frac{\omega}{T} \right) \phi_{sc}\left( \frac{\omega_{0,T}}{T} \right)
\eeq
Although this seemingly goes to zero at $\omega \to 0$, for $\omega \ll T$, we have $\coth(\beta \omega/2) \approx 2T/\omega$, which cancels with the $\omega/T$ factor and gives a non-zero limit.
This implies that in the very long-wavelength and low-frequency limit, the scaling of $T_1$ is set by
\beq
\frac{1}{T_1} \propto T^{(-2 + \eta)/z} \phi_{sc}\left( \frac{\omega_{0,T}}{T} \right), \text{ where } \omega_{0,T} \sim |\lambda - \lambda_c|^{z \nu}
\eeq
Therefore, by performing a scaling collapse of $1/T_1$ as a function of $T$ and $\lambda$, we can figure out both $(2 - \eta)/z$ and $z \nu$.
Therefore, all three critical exponents $\{\eta, \nu, z \}$ can be accessed via performing scaling collapses of the $1/T_1$ data in different regimes of operation of the qubit.

\section{Additional details for the Dipolaron Mode} \label{sec:dipolaron}
In this section, we provide the derivation of the dispersion of dipolarons in a two-dimensional fluid of electrically neutral dipolar molecules, and compute its contribution to dynamical polarization correlations. 

\subsection{Derivation of the dipolaron dispersion relation}

Like plasmons in a charged Fermi liquid, dipolarons are longitudinal collective modes that arise due to long-range electrostatic interactions in dipolar fluids.
We know that nature of plasmons in a Fermi liquid differ drastically between two and three spatial dimensions --- in $d=3$, plasmons are gapped excitations at $\q = 0$, while in $d=2$, they are gapless with a dispersion $\omega_p(\q) \propto \sqrt{q}$. 
The reason is the weaker electric field created by two dimensional charge imbalance results in a weaker restoring force at large distances, compared to a three dimensional charge imbalance.
Such an effect is at play for dipolarons too, resulting in gapless dispersion $\omega_{d}^2(\q) \sim a q^2 + b q^3$ for dipolarons in two dimensions.  
In what follows, we derive this dispersion from a simple hydrodynamic treatment of dipolar density fluctuations.
We note that our results are in accordance with more a microscopic treatment of collective modes in two-dimensional dipolar gases \cite{LiHwangDasSarma}. 
 
Consider a fluid of dipolar molecules at equilibrium density $n_d$ at chemical potential $\mu_{eq}$ and equilibrium velocity $\v_0 = 0$. Now we consider fluctuations about the mean density so that there is local density profile $\delta n(\r,t) = n(\r,t) - n_d$, and velocity $\v(\r,t) \neq 0$. The linearized continuity equation and Euler's (force) equation read as f$o$llows respectively:
\beq
\partial_t \delta n(\r,t) + n_d \nabla \cdot \v = 0, ~~ m n_d \partial_t \v = - n_d \nabla \left( \mu_{eq} + \frac{\partial \mu}{\partial n} \delta n \right) - \nabla (\bm{\mu} n_d \cdot \E)
\label{eq:Linearized}
\eeq
The generated electric field can be related to the fluctuating polarization density $\bm{\mu} n(\r)$ as discussed previously (neglecting retardation effects):
\beq
E_i(\r, t) &=& \int d\rp T^d_{ij}(\r, \rp) \mu_j n(\rp, t), ~~ T^d_{ij}(\q) = \frac{1}{4\pi \epsilon_0} \int d^dr e^{i \q \cdot \r} \partial_i \partial_j \left( \frac{1}{r} \right) = \begin{cases} -\frac{q_i q_j}{\epsilon_0 q^2}, ~ D = 3 \\ -\frac{q_i q_j}{2 \epsilon_0 q}, ~ D = 2 \end{cases}
\label{eq:Td}
\eeq
Going to momentum space and using that the isothermal compressibility is given by $\kappa = \frac{1}{n_d^2} (\frac{\partial n}{\partial \mu})$, we can combine Eqs.~(\ref{eq:Linearized}) to find the following equation for $\delta n(\q,\omega)$:
\beq
\omega^2 \delta n(\q,\omega) = \left( \frac{1}{\kappa m n_d} \q^2 + \frac{\mu_i \mu_j n_d}{m} \q^2 T^d_{ij}(\q) \right) \delta n(\q,\omega) 
\eeq
Using the form of $T_d$ from Eq.~(\ref{eq:Td}) for $D = 2$, we finally get the collective mode dispersion in $D=2$ that was quoted in the main text:
\beq
\omega_d^2(\q) = v^2 q^2 + \frac{n_d q (\q \cdot \bm{\mu})^2}{2 \epsilon_0 m} 
\eeq
where $v = \sqrt{1/\kappa m n_d}$ is the speed of the collective mode at small $\q$. This is analogous to the linearly dispersing zero sound mode in Fermi liquids, and does not require dipolar interactions. At larger momentum, anisotropy effects due to dipolar interactions come into play and we have a dominating $q^{3/2}$ term in the dispersion. In particular, if we take the angular average over all directions of $\bm{\mu}$ (which can point anywhere on the 2-sphere), then we can replace $(\q \cdot \bm{\mu})^2 \to q^2 \mu^2 /3$, and we recover the dispersion in Ref.~\onlinecite{LiHwangDasSarma} (converted to SI units). 
\beq
\omega_d^2(\q) = c_s^2 q^2 + \frac{\mu^2 n_d q^3}{6 \epsilon_0 m}
\eeq

\subsection{Contribution of dipolarons to the dynamical susceptibility}
Here, we derive the susceptibility $\chi(\omega,\q)$ that was used to calculate the dynamical polarization correlations for the neutral dipolar fluid. 
To this end, we closely follow Ref.~\onlinecite{Chaikin}, where the dynamical susceptibility $\chi(\omega,\q)$ is related to the Green's function $\bar{G}(\omega,\q)$ for the classical equation of motion for density fluctuations as
\begin{equation}
\bar{G}(\omega,\q) = \frac{1}{i \omega} \left( \frac{\chi(\omega,\q)}{\chi(\q)} -1 \right)
\label{eq:suppChiG}
\end{equation}
where $\chi(\q) = \chi(\omega = 0, \q)$ is the static susceptibility which approaches the compressibility $\chi_0$ in the $\q \to 0$ limit. 
Hence, we need to solve for the Green's function $\bar{G}(\omega,\q)$ for propagation of density fluctuations, defined as 
\begin{equation}
\bar{G}(\omega,\q) = \int_0^{\infty} dt \, e^{i \omega t} \int \frac{d^2q}{(2\pi)^2} \, e^{-i \q \cdot \r} \bar{G}(\r,t),  
\end{equation}
where $\bar{G}(\r,t)$ satisfies the following equation of motion and boundary condition:
\begin{equation}
\delta n(\r, t) = \int d^2 r^\prime \bar{G}(\r - \rp, t - t^\prime) \delta n(\rp, t^\prime), ~~~ \bar{G}(\r - \rp, t = 0) = \delta(\r - \rp) \implies \bar{G}(\q,t = 0) = 1
\end{equation}
Here, we focus on low-energy long-wavelength fluctuations arising from dipolarons, implying that $\bar{G}(\q,t)$ satisfies 
\begin{equation}
 \frac{\partial \bar{G}(\q,t)}{\partial t^2} + \omega_d^2(\q) \bar{G}(\q,t) = 0
\end{equation}
Together with the boundary condition $\bar{G}(\q,t = 0) = 1$, the solution to the above equation is given by:
\begin{equation}
\bar{G}(\q,t) = \cos(\omega_d(\q)t) \implies \bar{G}(\omega,\q) = \frac{i \omega}{\omega^2- \omega_d^2(\q)}
\end{equation}
As expected, the Green's function for propagation of density fluctuations has poles at the $\omega = \omega_d(\q)$ corresponding to undamped collective dipolaron modes. Correspondingly, using Eq.~\eqref{eq:suppChiG} the dynamical susceptibility is given by
\begin{equation}
\chi(\omega,\q) = \frac{\chi(\q) \, \omega_d^2(\q)}{ \omega_d^2(\q) - \omega^2} \xrightarrow[]{\q \to 0} \frac{\chi_0 \, \omega_d^2(\q)}{\omega_d^2(\q) - \omega^2}
\end{equation}
which is used in the main text for deriving the contribution of dipolarons to $1/T_1$.

\section{Derivation of Polarization Correlations for Relaxor Ferroelectric Model} \label{sec:relaxor}
In this section, we provide further details for the derivation of $1/T_1$ for a simple model of relaxor ferroelectrics. 
Specifically, we consider classical dipoles with slow relaxational dynamics, embedded in a material with dielectric constant $\varepsilon$. 
These dipoles are assumed to be correlated over a length-scale $\xi$, which may be interpreted as the size of a typical polar nano-region which is often used to model relaxor ferroelectrics. 

The field due to a single electrical dipole with dipole moment $\p$ at the probe location $(0,0,d)$, located at $(x,y,-z)$ is (neglecting retardation effects):
\beq
\E(\r,t) = -\frac{1}{4 \pi \varepsilon_0} \left( \frac{2}{\varepsilon_p + \varepsilon} \right) \left( \frac{\p(t) - 3 \hat{r} (\hat{r} \cdot \p(t))}{r^3} \right)
\eeq
where $\r = (-x, -y, d+z)$, $\hat{r} = \r/|\r|$ denotes the unit vector along $\r$, and the factor of $ \left( \frac{2}{\varepsilon_p + \varepsilon} \right)$ comes from solving the image charge problem.
Therefore, the total electric field due to all such dipoles can be obtained using the principle of superposition by summing up the fields from all dipoles:
\beq
\E(t) = -\sum_\r \frac{1}{4 \pi \varepsilon_0} \left( \frac{2}{\varepsilon_p + \varepsilon} \right) \left( \frac{\p(\r,t) - 3 \hat{r} (\hat{r} \cdot \p(\r,t))}{r^3} \right)
\eeq
Let us assume that the dipoles are uniformly distributed with density $n = 1/\ell^3$ in the dielectric.
If the qubit-sample distance is much larger than the average separation between the dipoles, $d \gg \ell$, then we can approximate the field by its continuum limit:
\beq
\E(t) = -\frac{1}{\ell^3}  \int_{\r} \frac{1}{4 \pi \varepsilon_0} \left( \frac{2}{\varepsilon_p + \varepsilon} \right) \left( \frac{\p(\r,t) - 3 \hat{r} (\hat{r} \cdot \p(\r,t))}{r^3} \right)
\eeq
where $\int_{\r} = \int_0^\infty dz \int_{-\infty}^\infty dx \int_{-\infty}^\infty dy$, assuming a semi-infinite sample.
Therefore, we may write the retarded electric field correlators, our central correlation functions of interest, in frequency space, as the averaged response:
\beq
\chi^R_{E_i,E_j}(\omega) = \left(\frac{1}{4 \pi \epsilon_0} \right)^2 \left( \frac{2}{\varepsilon_p + \varepsilon} \right)^2 \frac{1}{\ell^6} \int_{\r} \int_{\rp} \frac{\langle p_i p_j \rangle 
- 3 \langle p_i p_\alpha \rangle \hat{r}^\prime_\alpha \hat{r}^\prime_j 
- 3 \langle p_\alpha p_j \rangle \hat{r}_i \hat{r}_\alpha + 9 \langle p_\alpha p_\beta \rangle \hat{r}_\alpha \hat{r^\prime}_\beta \hat{r}_i \hat{r^\prime}_j}{r^3 (r^\prime)^3}
\label{eq:chiE}
\eeq
where $\langle p_i p_j \rangle$ is shorthand for $\int_0^\infty dt \, e^{i(\omega + i 0^+)t} \langle p_i(\r,t) p_j(\rp,0) \rangle$. 

If we assume that the dipoles are isotropic, then the most general expression one may write down for the correlation is:
\beq
\langle p_i(\r,t) p_j(\rp,0) \rangle = \frac{|\p|^2}{3} \delta_{ij}  C(\r, \rp, t), \text{ where } C(\r, \r, 0) = 1
\eeq
In presence of translation invariance (assuming the qubit is reasonably far for the sample), we have $C(\r, \rp, t) = C(\r - \rp, t)$. 
We may further assume factorizability of correlations in space and time for simplicity
\beq
C(\r, \rp, t) = C(\r - \rp, t) = C(\r - \rp) f(t)
\eeq
Depending on the nature of the system, we may have exponential or power law decay in spatial correlations. 
Further, for simple Debye relaxational dynamics, $f(t) = e^{-t/\tau}$. 

In the extreme limit of uncorrelated dipoles, we can make the following substitution:
\beq
C(\r - \rp) = \delta_{\r, \rp} \to \ell^3 \, \delta(\r - \rp) \text{ in the continuum limit}
\eeq 
In this analytically tractable limit, the retarded correlator in Eq.~\eqref{eq:chiE} takes the following form:
\beq
\chi^R_{E_i,E_j}(\omega) = \left(\frac{1}{4 \pi \epsilon_0} \right)^2 \left( \frac{2}{\varepsilon_p + \varepsilon} \right)^2 \frac{1}{\ell^3} \int_{\r}  \frac{\langle p_i p_j \rangle - 3 \langle p_i p_\alpha \rangle \hat{r}_\alpha \hat{r}_j - 3 \langle p_\alpha p_j \rangle \hat{r}_i \hat{r}_\alpha + 9\langle p_\alpha p_\beta \rangle \hat{r}_\alpha \hat{r}_\beta \hat{r}_i \hat{r}_j}{r^6}
\eeq
Further, using $\langle p_i p_j \rangle = \frac{1}{3} |\p^2| \tilde{f}(\omega) \delta_{ij}$ and $|\hat{r}|^2 = \hat{r}_\alpha \hat{r}_\beta \delta_{\alpha \beta} = 1$, we have
\beq
\chi^R_{E_i,E_j}(\omega) = \left(\frac{1}{4 \pi \epsilon_0} \right)^2 \left( \frac{2}{\varepsilon_p + \varepsilon} \right)^2 \frac{1}{3} |\p^2| \tilde{f}(\omega)  \frac{1}{\ell^3} \int_{\r} \frac{\delta_{ij} + 3 \hat{r}_i \hat{r}_j}{r^6}
\eeq
where we have defined $\tilde{f}(\omega) = \int_0^\infty dt \, e^{i \omega t} f(t)$.
Using reflection symmetry about the xz and yz planes, one can show that the off diagonal terms vanish after spatial integration (i.e., $\int_\r \hat{r}_i \hat{r}_j/r^6 = 0$ for $i \neq j$), so we have
\beq
\chi^R_{E_i,E_j}(\omega) = \left(\frac{1}{4 \pi \epsilon_0} \right)^2 \left( \frac{2}{\varepsilon_p + \varepsilon} \right)^2 \frac{1}{3} |\p^2| \tilde{f}(\omega) \delta_{ij} \left[ \frac{1}{\ell^3} \int_{\r} \frac{1 + 3 \hat{r}_i^2}{r^6} \right]
\eeq
We can use the following integrals (recall that the finite sample-qubit distance $d$ acts as a natural cutoff that avoids divergence in the integrals):
\beq
\int_\r \frac{1}{r^6} = \frac{\pi}{6 d^3}, ~~~ \int_\r \frac{\hat{r}_i^2}{r^6} = \begin{cases}
    \frac{\pi}{9 d^3}, ~ i = z \\
    \frac{\pi}{36 d^3}, ~ i = x,y
\end{cases}
\eeq
Therefore, we finally have
\beq
\chi^R_{E_i,E_j}(\omega) = \left(\frac{1}{4 \pi \epsilon_0} \right)^2 \left( \frac{2}{\varepsilon_p + \varepsilon} \right)^2 \frac{1}{3 \ell^3} |\p^2| \tilde{f}(\omega) \delta_{ij} \left[ \frac{\pi}{2d^3} \delta_{i,z} + \frac{\pi}{4d^3} (\delta_{i,x} + \delta_{i,y}) \right]
\eeq
For the simple case of Debye relaxational dynamics, we have 
\beq
\tilde{f}(\omega) = \int_0^\infty dt \, e^{i (\omega + i 0^+)t} e^{-t/\tau} =  \frac{1}{- i \omega + 1/\tau}
\eeq
Therefore, the lifetime is given by
\beq
\frac{1}{T_1} = n \, d_\perp^2 \coth(\beta \omega) \text{Im}[\chi^R_{E_+,E_-}(\omega)] = n \, d_\perp^2 \left( \frac{\coth(\beta \omega) \omega \tau^2}{1 + \omega^2 \tau^2}  \right) \left(\frac{1}{4 \pi \epsilon_0} \right)^2 \left( \frac{2}{\varepsilon_p + \varepsilon} \right)^2 \frac{|\p^2|}{4 d^3} 
\label{eq:T1uncorr}
\eeq
where we have oriented the qubit in the x-y plane to maximize the response, and used that $n = 1/\ell^3$ is the density of dipoles. 
Note that when $\beta \omega \ll 1$ (regime where NVs typically operate), then $\omega \coth(\beta \omega) \approx k_B T$, so the noise is maximized at $\omega = 0$.
Conversely, when $\beta \omega \gg 1$, then $\coth(\beta \omega) \approx 1$ and the noise is maximized at $\omega \approx \tau^{-1}$.

Now we can generalize this result to the scenario where the spatial correlations between dipoles are not `ultra short-ranged'. 
To this end, we consider two possible spatial correlation profiles:
\beq
C(\r- \rp) = e^{-|\r - \rp|/\xi}, \text{ and } C(\r- \rp) = |\r - \rp|^{-\alpha} \text{ for } |\r - \rp| \gg \text{ lattice spacing}
\eeq
Using identical symmetry arguments as above, we find that
\beq
\begin{gathered}
\chi^R_{E_i,E_j}(\omega) = 
\left(\frac{1}{4 \pi \epsilon_0} \right)^2 \left( \frac{2}{\varepsilon_p + \varepsilon} \right)^2 
\frac{1}{3} |\p^2| \tilde{f}(\omega) I_{ij}(\xi,d),
\\
I(\xi,d)
=
\frac{1}{\ell^6} 
\int_{\r} \int_{\rp} \frac{\delta_{ij} - 3 \hat{r}_i \hat{r}_j 
- 3 \hat{r^\prime}_i \hat{r^\prime}_j
+ 9 \sum_\alpha \hat{r}_\alpha \hat{r^\prime}_\alpha \hat{r}_i \hat{r^\prime}_j}{r^3 (r^\prime)^3}
C(\r - \rp)
\end{gathered}
\eeq

While an analytic evaluation of the spatial integrals are no longer possible, we can still extract useful information via scaling. 
In case of exponential correlations, when $\xi \ll d$, then the exponential constrains $|\r - \rp| \lesssim \xi$, implying that the $\rp$ integral just yields an additional factor of $\xi^3$.
In this regime, $\frac{1}{T_1} \propto n^2 \xi^3/d^3$ (i.e., our result in Eq.~\eqref{eq:T1uncorr} gets multiplied by $n \xi^3$). 
This makes physical sense, as the typical size of each dipolar region is now of $O(\xi^3)$, so it behaves as a single dipole.
The number of such correlated dipolar regions is given by $n \xi^3$, and since these dipolar regions fluctuate independently, their contributions to the electrical noise add up, leading to the additional factor of $n \xi^3$ in $1/T_1$.

In the other limit when $\xi \gg d$, we can carry out the $\r$ and $\rp$ integrals independently to lengthscales of $\xi$, where the exponential factor is equal to 1. Neglecting the contributions of $r,r'\geq \xi$ (which converges at large $r,r'$ due to decaying $C(\r-\rp)$), we can approximate the result by:

\beq
\begin{gathered}
I(\xi,d)
\approx_{\xi\gg d}
\frac{1}{\ell^6} 
\int_{\r,r<\xi}  
\int_{\rp,r'<\xi} 
\frac{\delta_{ij} - 6 \hat{r}_i \hat{r^\prime}_j + 9 \sum_\alpha \hat{r}_\alpha \hat{r^\prime}_\alpha \hat{r}_i \hat{r^\prime}_j}{r^3 (r^\prime)^3}=
\\
=
\left\langle
\left(\int_{\r,r<\xi}  
\frac{n_i - 3 \hat{r} (\hat{r} \cdot {\bf n})}{r^3}
\right)
\left(
\int_{\rp,r'<\xi}  
\frac{n_j - 3 \hat{r}^\prime (\hat{r}^\prime \cdot {\bf n})}{(r^\prime)^3}
\right)
\right\rangle_{\bf n},
\end{gathered}
\label{eq:sup:xigd}
\eeq
where $\langle n_i n_j\rangle = \delta_{ij}$, such that the second line is identical to the first one. Each integral inside the brackets in Eq. \eqref{eq:sup:xigd} represents a field of a dielectric region of linear size $\xi$ with polarization density ${\bf n}$. In principle, the field depends on the shape of the region, but for a region of a rather generic ellipsoidal shape the field at the surface of the ellipsoid is simply proportional to ${\bf n}$, i. e. $E_i = \sum_j \alpha_{ij} n_j$ \cite{clark1945}, with the proportionality coefficient being a number dependent on the exact shape. Far away from the surface of the region, at a distance $R \gg \xi$, one expects the field to decay in a dipolar way as $\sim n \xi^3/R^3$. For the case $d\ll \xi$, the former case is more relevant, resulting in:
\beq
I_{ij}(\xi,d)\big|_{\xi\gg d}
\approx
\sum_{i',j'}\alpha_{ii'} \alpha_{ij'} \langle n_{i'}n_{j'}\rangle
=
\sum_{i'} \alpha_{i,i'}^2,
\eeq
i.e., in the case $d\ll\xi$, $\chi^R_{E_i,E_j}(\omega) $ and with it, $1/T_1$ simply saturates to a constant, $d$-independent value.
The key take-away is that a crossover from $1/d^3$ scaling to an approximately $d$-independent scaling of $1/T_1$ can help us estimate the size of a polar nano-region, as discussed in the main text.

As an aside, our formalism also allows us to evaluate the $d$ scaling of $\chi^R_{E_i,E_j}(\omega)$ in case of power-law spatial correlations - corresponding to a critical disordered phase. 
One again, there is no analytical solution, but since the only length-scale in the $\r, \rp$ integrals is $d$, $\chi^R_{E_i,E_j}(\omega)$ must scale as $d^{-\alpha}$ when $C(\r- \rp) = |\r - \rp|^{-\alpha}$ at long distances.

\end{document}